\documentclass[12pt,a4paper]{article}

\usepackage[utf8]{inputenc}
\usepackage[T1]{fontenc}
\usepackage{lmodern}
\usepackage{geometry}
\usepackage{times}      
\usepackage{setspace}
\usepackage{amsmath, amsfonts, amssymb, bm, dsfont, relsize, exscale, bigints}

\usepackage{graphicx}
\usepackage{booktabs}
\usepackage{multirow}
\usepackage{tabularx}
\usepackage{array}
\usepackage{makecell}
\usepackage{arydshln}
\usepackage{dcolumn}
\usepackage{float} 
\usepackage{caption}
\usepackage{placeins}  

\usepackage[superscript,biblabel]{cite} 
\makeatletter
\def\@biblabel#1{#1.} 
\makeatother

\usepackage{xcolor}
\usepackage[colorlinks=true,linkcolor=blue,citecolor=blue,urlcolor=teal]{hyperref}
\usepackage{xurl}

\usepackage{fancyhdr}
\usepackage{url}
\usepackage{appendix}
\usepackage{chngcntr}

\newcommand{\HRule}{\rule{\linewidth}{0.5mm}}

\title{\HRule \\[0.4cm]
{\LARGE \bfseries A Functional Approach to Testing Overall Effect of Interaction Between DNA Methylation and SNPs}\\[0.4cm]
\HRule}

\author{
Yvelin GANSOU\\
\parbox[t]{0.9\textwidth}{\centering Département de Mathématiques et de Statistique, Université Laval, Québec, QC, Canada}\\
\texttt{yvelin.gansou.1@ulaval.ca}
\and
Karim Oualkacha\\
\parbox[t]{0.9\textwidth}{\centering Département de Mathématiques, Université du Québec à Montréal, Montréal, QC, Canada}\\
\texttt{oualkacha.karim@uqam.ca}
\and
Marzia Angela Cremona\\
\parbox[t]{0.9\textwidth}{\centering Chair in statistical learning and Département d’opérations et systèmes de décision, Université Laval, Québec, QC, Canada}\\
\texttt{marzia.cremona@fsa.ulaval.ca}
\and
Lajmi Lakhal-Chaieb\\
\parbox[t]{0.9\textwidth}{\centering Département de Mathématiques et de Statistique, Université Laval, Québec, QC, Canada}\\
\texttt{lajmi.lakhal@mat.ulaval.ca}
}
\date{}

\begin{document}
\maketitle

\vspace{1cm}

\begin{abstract}
 We introduce a test for the overall effect of interaction between DNA methylation and a set of single nucleotide polymorphisms (SNPs) on a quantitative phenotype. The developed inference procedure is based on a functional approach that extents existing regression models in functional data analysis. Through extensive simulations, we show that the proposed test effectively controls type I error rates and highlights increased empirical power over existing methods, particularly when multiple interactions are present. The use of the proposed test is illustrated with an application to data from obesity patients and controls.
\end{abstract}


\vspace{1cm}


\maketitle

\section{Introduction}
\label{sec1}

\noindent Epigenetics is the study of reversible and heritable changes in gene expression that occur without altering the DNA sequence itself. It plays a crucial role in understanding complex diseases and disorders, as well as exploring the links between environmental factors and diseases \cite{jirtle2007environmental}.
DNA methylation is the most studied epigenetic process. It usually occurs at specific locations in the DNA sequence called CpGs sites. The human genome contains approximately 28.2 million CpG sites \cite{khodasevich2022comparison}. DNA methylation is involved in the regulation of gene expression, in tissue differentiation and in the development of cancer \cite{hansen2011generalized}. \\
DNA methylation data takes different forms depending on the technology used to generate the measurements (see reference \cite{khodasevich2022comparison} for a comparison between existing technologies). The Illumina HumanMethylationEPIC BeadChip (EPIC) DNA microarray is the most widely employed technology and it is able to measure DNA methylation at $m \approx 935 000$ pre-selected fixed CpG sites throughout the human genome \cite{guanzon2024comparing}. Data generated by this technology for $N$ independent individuals can be presented as $\{(t_j,p_{ij}), i=1,\cdots,N; j=1,\cdots,m\}$, where $p_{ij} \in [0,1]$ is a raw estimate of the DNA methylation level for the $j^{th}$ CpG site located at the genomic position $t_j$. These estimates are computed as the ratio of the number of methylated nucleotides to the total number of nucleotides at each covered CpG site. Recently, targeted Next-Generation Sequencing (NGS)-based technologies have been developed to obtain DNA methylation measures at a single base resolution, which provides a better coverage of the human genome. With this technology, the genomic positions and the total number of covered CpG sites varies from an individual to another. Therefore, the resulting data takes the form $\{(t_{ij},p_{ij}), i=1,\cdots,N; j=1,\cdots,m_i\}$. In practice, $m_i$ varies from 4 to 28.2 millions \cite{guanzon2024comparing}.\\
The DNA methylation levels have a spatial correlation structure, as measures at neighbouring genomic positions tend to have close values. Therefore, $p_{ij}$ can be seen as an estimate of the true DNA methylation level $\Pi_i(t_{ij})$, where $\Pi_i(\cdot)$ is a smooth, continuous curve. 
Several researchers have developed statistical methods to identify genomic regions where the DNA methylation profile is significantly associated with phenotypes \cite{hansen2012bsmooth,lakhal2017smoothed,milad2017functional,zhao2020novel}. DNA methylation profiles in specific genomic regions has been shown to characterize many diseases, such as diabetes \cite{nilsson2014altered}, Alzheimer's disease \cite{de2014alzheimer} and certain autoimmune diseases \cite{liu2013epigenome}. \\
SNPs (Single Nucleotide Polymorphisms) are very common genetic variations in humans. They are important for understanding genetic variability in populations and  identifying associations between genetic variations and diseases or physical characteristics. 
SNPs data for individual $i$ can be represented as $\{G_{id}, d=1,\cdots,D\}$, where $G_{id} \in \{0,1,2\}$ is the count of the minor allele of the $d^{th}$ SNP located at the genomic position $u_d$. In practice, the total number of SNPs $D$ involved in an analysis can be smaller than 10 for some specific settings and can reach several hundreds of thousands in genome-wide association studies (GWAS).
As demonstrated by previous studies, some SNPs are associated with the development of diseases such as cancer, cardiovascular diseases and obesity \cite {easton2008genome,zeller2012genomewide, willer2009six}. \\
Traditionally, researchers have primarily focused on the effects of DNA methylation and SNPs on phenotypic outcomes separately. Recently, there has been a growing interest in understanding the interplay between epigenetic modifications and genetic variation, particularly how DNA methylation and SNPs jointly influence traits. This shift in focus acknowledges that epigenetic and genetic factors do not operate in isolation but interact in complex ways to affect gene expression and phenotypic outcomes. For instance, a previous study \cite{soto2013interaction} shows that DNA methylation in the interleukin-4 receptor is associated with asthma, but this association is further explained by the presence or absence of a nearby SNP. 
More recently, Veenstra et al.\cite{veenstra2018epigenome} and Wang et al.\cite{wang2022genetic} conducted studies to investigate the effect of the interaction between DNA methylation and SNPs on triglyceride levels and newborn telomere length, respectively, using DNA methylation data obtained with Illumina technology.
 To this end, they selected a large number of pairs (CpG site $j$, SNP $d$) located at neighbouring genomic positions; i.e., requiring $|t_j-u_d|$ smaller than a predetermined threshold. For each selected pair, they fitted a separate linear model and tested the significance of the interaction term $p_{ij} \times G_{id}$. Finally, they adjusted the significance threshold to account for the large number of tests performed, using multiple testing correction methods such as Bonferroni or FDR (False Discovery Rate) correction. 
The above described ad-hoc procedure has two main drawbacks. First, it requires performing a very large number of tests, which makes the control of type 1 error rate a complex task. For example, Veenstra et al.\cite{veenstra2018epigenome} selected approximately 700,000 (CpG site $j$–SNP $d$) pairs  and considered two adjusted significance thresholds obtained by a rule of thumb: $5 \times 10^{-8}$ and $1\times 10^{-4}$; in such a setting, it is not clear how the type 1 error rate is controlled. Second, this methodology totally ignores the spatial correlation structure of DNA methylation levels, which may result in a significant loss of power. \\
In this work, we consider a functional data analysis approach to test the significance of the overall effect of interaction between DNA methylation levels and a set of SNPs on a continuous phenotype. The proposed approach avoids multiple testing and takes into account the spatial correlation structure of DNA methylation levels. From a statistical point of view, we represent the DNA methylation data as a smooth continuous curve $\{\Pi(t), t \in {\cal G}\}$ over the genomic region ${\cal G}$, whereas the SNPs data consist of $D$ scalar variables $G_d$ with genomic positions $\{u_1,\cdots,u_D\} \in {\cal G}^D$. The presence of a curve predictor prompts us to consider a functional linear model with a scalar response (i.e., a scalar-on-function model; see Ramsay and Silverman).\cite{silverman2005} This type of model was first introduced by Ramsay and Dalzel \cite{ramsay1991some}  in the case of a single functional covariate and then extended to include additive effects of multiple functional predictors (see, e.g., Goldsmith et al \cite{goldsmith2011penalized}; Ivanescu et al., \cite{ivanescu2015penalized}). In their work, Usset et al. \cite{usset2016interaction} introduced an interaction term between a pair of functional predictors in scalar-on-function regression models, while Beyaztas and Shang \cite{beyaztas2021partial} considered a similar interaction term between pairs of functional predictors in the case of function-on-function regression. Finally, Meyer et al. \cite{meyer2015bayesian} considered interaction terms between scalars and functional predictors in Bayesian function-on-function mixed models. 
In this work, we consider $D$ interaction terms between the functional covariate $\{\Pi(t), t \in {\cal G}\}$ and a set of scalar predictors $G_d$ located at specific positions $\{u_1,\cdots,u_D\} \in {\cal G}^D$, within the same metric space. Differently from Meyer et al. \cite{meyer2015bayesian}, which only considers the product between the scalar and the functional covariates as interaction terms, we introduce a non-increasing weight function in $|t-u_d|$ to express the interaction terms. This allows us to consider the entire methylation curve $\Pi(t)$, while putting more weight on the locations close to the SNPs and less weight on the ones farther away. Finally, we derive an inference procedure to test the significance of these interactions. \\
The article is organised as follows. Section \ref{sec2} describes the proposed model along with estimation and testing procedure of the interaction. Simulation studies evaluating the performance of our method are presented  in Section \ref{sec3}. In Section \ref{sec4}, we apply the proposed model to real data to study the interaction between obesity-associated SNPs and DNA methylation measures at CpG sites and compare the performance of our method to existing methods in term of type I errors and power. The paper concludes with a discussion in Section \ref{sec5}.

\newpage
\section{Methods}
\label{sec2}

\subsection{Notation and data}
\label{sec2.1}

Suppose we have DNA methylation measurements $\{(t_{ij},p_{ij}), i=1,\cdots,N; j=1,\cdots,m_i\}$ and minor alleles counts data $\{G_{id} \in \{0,1,2\}, i=1,\cdots,N; d=1,\cdots,D\}$ of $D$ SNPs located at $\{u_1,\cdots,u_D\}$  in a given genomic region for $N$ independent individuals.

\noindent
We transform methylation data into  functional form by applying kernel smoothing methods, following the approach of Hansen et al. \cite{hansen2012bsmooth}. This allows us to estimate methylation levels at any position $t$ within the genomic region $[ \min (t_{ij}),\max(t_{ij})]$. Estimation at any 
$t$ results in the construction of a methylation curve for the individual $i$, denoted as $\Pi_i(t)$. 
This is given by :
\begin{eqnarray}\label{eq21.1}
     \Pi_i(t) = \frac{\displaystyle \sum_{j=1}^{m_i}K \left(\frac{t_{ij}-t}{h}\right) p_{ij}  }{\displaystyle\sum_{j=1}^{m_i} K \left(\frac{t_{ij}-t}{h} \right) } , 
\end{eqnarray}
where $K$ is the density function of the standard normal distribution and $h$ is a bandwidth parameter that controls the smoothness of the procedure. There is a large literature on ways to select the bandwidth $h$; see this work \cite{doksum2000variable} and the references
therein for a comprehensive review. In this context, the selection of $h$ determines how locally smoothed are the
methylation levels. Here, we follow this approach \cite{ramachandran2013adaptive}  and consider an adaptive bandwidth $h$ that varies with the genomic position $t$. A popular approach in genomics sets $h = \max \{d_k(t), h_{min} \} $, where
$d_k(t)$ is the distance from $t$ to the $k^{th}$ nearest CpG site and $h_{min}$ is the minimum value that $h$ can take \cite{hansen2012bsmooth}. We set $k = 70$ and $h_{min} = 1$ kb, which are the default values used by the Bioconductor package BSmooth \cite{hansen2012bsmooth}.  Without loss of generality, we scale  $ [ \min (t_{ij}),\max(t_{ij})] $ to $[0,1]$ in the model.

\noindent Additionaly, suppose that for $i=1, \ldots, N $, we observe a continuous response variable $Y_i$ and $S$ non-genetic covariates $W_i = (W_{i1},..., W_{iS}) $. Our objective is to test the significance of the interaction between SNPs and DNA methylation.

\subsection{Previous work}

Veenstra et al. \cite{veenstra2018epigenome} and Wang et al. \cite{ wang2022genetic} considered DNA methylation data obtained with Illumina technology $\{(t_j,p_{ij}), i=1,\cdots,N; j=1,\cdots, m\}$. They fixed a threshold $l$ and selected all pairs of CpG site $j$ and SNP $d$ satisfying $|t_j-u_d|<l$. For each selected pair $(j,d)$, they fitted the linear model

 \begin{eqnarray}
 \label{Wang_model}
    Y_i =\zeta_0 + \sum_{s=1}^S\zeta_s W_{is} + \alpha G_{id} +    \delta p_{ij} +  \gamma G_{id} p_{ij} + \varepsilon_i ,
   \end{eqnarray}
where $\zeta_0$ is the model intercept, while $\zeta_s$, $\alpha$ and $\delta$ are fixed-effect parameters of the $s^{th}$ covariate, the $d^{th}$ SNP and the $j^{th}$ CpG site, respectively.  The parameter $\gamma$ denotes the effect of the CpG-SNP interaction. The authors computed the $p$-value of the test $H_0: \gamma=0$ versus $H_1: \gamma \neq 0$ and identified an interaction between $G_{id}$ and $p_{ij}$ if the obtained $p$-value is smaller than a global adjusted significance threshold. \\
Of note, if one assumes a hard-threshold function of the $L_1$ distance between a SNP at position $u$ and a CpG site at position $t$, defined as $\varphi(t,u) = \mathbf{1}\{|t-u|<l\}$, then the interaction term in (\ref{Wang_model}) can be expressed as $\gamma \varphi(t_{ij},u_{id}) G_{id} p_{ij}$. The key innovation of our proposed approach is to move beyond this simple thresholding by incorporating both smooth functions of $L_1$ distance of neighboring SNP–CpG pairs and the complete methylation curve, thereby enabling a more flexible and effective modeling of such interactions, as described next.

\subsection{Model}
\label{sec2.2}

In this work, we consider the functional linear model

 \begin{eqnarray}
    Y_i =\zeta_0 + \sum_{s=1}^S\zeta_s W_{is} + \sum_{d=1}^D \alpha_d G_{id} +    \int_0^1 \delta(t) \Pi_i(t) dt + \sum_{d=1}^D   G_{id} \int_0^1 \varphi_d (  t , u_d  ) \Pi_i(t) dt + \varepsilon_i ,
   \end{eqnarray}
where  $\zeta_0 $ is the intercept,
       $\{ \zeta_s \}_{s=1}^S$ are the coefficients associated to confounding variables effects, $\{ \alpha_d \}_{d=1}^D$ are the coefficients associated to SNPs effects,
     $\delta(t)$ is a functional coefficient associated to DNA methylation effect and
      $\varphi_d ( t , u_d ) $ is a bivariate functional coefficient defined on $[0,1] \times [0,1]$ associated to the interaction effect between the $d^{th}$ SNP $G_{id}$ located at genomic position $u_d$ and DNA methylation  $ \Pi_i(t) $. Finally, the error terms $\{\varepsilon_i, i=1,\cdots,N\}$ are assumed to be independent and identically distributed following $N(0, \sigma^2)$.
      
\noindent We write $\varphi_d$ as
\begin{eqnarray}
  \varphi_d (t,u_d) = \eta_d \psi_{\rho} (\vert t - u_d \vert),
\end{eqnarray}
where $\eta_d$ is a regression coefficient, and $\psi_{\rho}: [0,1] \rightarrow [0,1]$ is a non-increasing weight function indexed by a positive parameter $\rho$, which controls the extent of the interaction that we want to test. For the sake of identifiability, we impose $\psi_{\rho}(0) = 1 $. 
The interaction parameters $\eta_d$ quantify how the local DNA methylation profile modulates the effect of SNP $d$ on the phenotype. In practice, they can be interpreted as the contribution of SNP–methylation interplay to the variance of $Y$, beyond their marginal effects.
In this work, we consider three specific parametric forms for  $\psi_{\rho}$: $\psi_{\rho}(u)= e^{-\rho u}$ (convex), \mbox{$\psi_{\rho}(u) = e^{- \rho^2 u^2 }$} (concave) and $\psi_{\rho}(u) = \max \{ 1 - \rho u, 0 \}$ (linear). Therefore, the resulting model is \begin{eqnarray}\label{eq22.3}
   Y_i =\zeta_0 + \sum_{s=1}^S\zeta_s W_{is} + \sum_{d=1}^D \alpha_d G_{id} +    \int_0^1 \delta(t) \Pi_i(t) dt + \sum_{d=1}^D \eta_d G_{id} \int_0^1 \psi_{\rho}( | t - u_d | ) \Pi_i(t) dt + \varepsilon_i. 
  \end{eqnarray}

\subsection{Estimation}
\label{sec2.3}

Our methodological framework falls within the class of scalar-on-function regression models (SoFR)\cite{silverman2005, crainiceanu2024functional}. We expand the coefficient function in a spline basis and impose a quadratic penalty on the second derivative, an approach introduced by O'Sullivan \cite{o1986statistical} and Wahba \cite{wahba1990spline}, and further developed by Reiss and Ogden \cite{reiss2009smoothing} and by Eilers and Marx \cite{eilers1996flexible}. This penalized spline representation is equivalent to a linear mixed model, enabling estimation via REML using existing software such as \texttt{mgcv}  \cite{goldsmith2011penalized, usset2016interaction,crainiceanu2024functional}.

\noindent In particular, we consider basis function expansion of the coefficient function $\delta(t)$ using B-Splines basis. Let $ \left\lbrace B_{\ell}(t) \right\rbrace_{\ell=1}^L $ be a B-Splines basis. We assume $ \delta(t) = \sum_{\ell=1}^L b_{\ell} B_{\ell}(t) ,$ where $b_{\ell}$ are the unknown parameters. Thus, estimation of
the coefficient function $\delta(t)$ is reduced to estimation of the unknown coefficients $b_{\ell}$. We have that

 $$   \int_0^1 \delta(t) \Pi_i(t) dt 
      = \sum_{\ell=1}^{L} b_{\ell} \left( \int_0^1 B_{\ell}(t) \Pi_i(t) dt \right) \\  
      = \sum_{\ell=1}^{L} b_{\ell} Z_{i\ell} ,  $$

$$     \sum_{d=1}^D \eta_d G_{id} \left( \int_0^1 \psi_{\rho}( | t - u_d |) \Pi_i(t) dt \right) = \sum_{ d=1}^D \eta_d G_{id} \Omega_{id,\rho} , $$
where  $ Z_{i\ell} = \int_0^1 B_{\ell}(t) \Pi_i(t) dt $ and $ \Omega_{id, \rho} = \int_0^1 \psi_{\rho}( | t - u_d |) \Pi_i(t) dt$.\\
Hence, we can write 
\begin{eqnarray}
    Y_i =\zeta_0 +  \boldsymbol{W}_i^{\top} \boldsymbol{\zeta}  + \boldsymbol{G}_i^{\top} \boldsymbol{\alpha}  + \boldsymbol{Z}_i^{\top} \boldsymbol{ b }   + \boldsymbol{K}_{i,\rho}^{\top} \boldsymbol{\eta}   + \varepsilon_i ,
   \end{eqnarray}
where we assume that  $ \boldsymbol{\zeta} = (\zeta_1,\ldots , \zeta_S)^{\top} $ , $ \boldsymbol{W_i}^{\top} = (W_{i1},\ldots , W_{iS})$ are vectors of length $S$;  $\boldsymbol{\alpha} = (\alpha_1,\ldots , \alpha_D)^{\top} $, $\boldsymbol{G_i}^{\top} = (G_{i1},\ldots , G_{iD})$,  $ \boldsymbol{\eta} = (\eta_1,\ldots , \eta_D)^{\top} $ and $\boldsymbol{K_{i,\rho}}^{\top} = (G_{i1}\Omega_{i1,\rho},\ldots , G_{iD}\Omega_{iD,\rho})$ are vectors of length $D$; 
$ \boldsymbol{ b} = ( b_1,\ldots ,  b_L)^{\top}$ and $\boldsymbol{Z_i}^{\top} = (Z_{i1},\ldots , Z_{iL})$ are vectors of length $L$.\\
By considering the matrices
$$ \boldsymbol{W} = \begin{pmatrix}
	W_{11} &  \ldots & W_{1S}  \\  
	\vdots &  \vdots & \vdots   \\ 
    W_{N1} &  \ldots & W_{NS} 
\end{pmatrix} ,  \quad    
\boldsymbol{G} = \begin{pmatrix}
	G_{11} &   \ldots & G_{1D}  \\  
	\vdots &  \vdots & \vdots   \\ 
	G_{N1} & \ldots & G_{ND} 
\end{pmatrix} ,$$

$$ \boldsymbol{Z} = \begin{pmatrix}
	Z_{11} &  \ldots & Z_{1L}  \\  
	\vdots &  \vdots & \vdots   \\ 
    Z_{N1} &  \ldots & Z_{NL} 
\end{pmatrix} ,    \qquad    
\boldsymbol{K} = \begin{pmatrix}
	G_{11}\Omega_{11} &   \ldots & G_{1D}\Omega_{1D}  \\  
	\vdots &  \vdots & \vdots   \\ 
	G_{N1}\Omega_{N1} & \ldots & G_{ND}\Omega_{ND} 
\end{pmatrix} ,$$
we obtain
\begin{eqnarray*}
    \boldsymbol{Y} = \boldsymbol{A} \boldsymbol{\theta}  + \boldsymbol{\varepsilon} ,
\end{eqnarray*}
where   $\boldsymbol{A}= (\mathds{1}_N | \boldsymbol{W}|\boldsymbol{G}|\boldsymbol{K}|\boldsymbol{Z} )$ is the overall model design matrix of dimension $N \times (S+2D+L+1) $; $ \boldsymbol{\theta} = ( \zeta_0,\zeta_1,\ldots,\zeta_S,\alpha_1,\ldots,\alpha_D, \eta_1,\ldots,\eta_D, b_1,\ldots, b_L)^{\top}$ is a vector of length $(S+2D+L+1)$ and $\boldsymbol{Y} = (Y_1, \ldots , Y_N)^{\top},  \boldsymbol{\varepsilon} = (\varepsilon_1,\ldots, \varepsilon_N)^{\top} $ are vectors of length $N$.

\noindent To estimate $\boldsymbol{\theta}$, one may proceed by ordinary least squares (OLS), which is equivalent to maximizing the log-likelihood function under the assumption of normally distributed errors.
To control the smoothness of the coefficient function $\delta(t)$, we add a roughness penalty to the log-likelihood function. The idea of roughness penalty on the integrated squared second derivative was first proposed by O'Sullivan \cite{o1986statistical} in the context of penalized splines, following earlier work on smoothing splines by Wahba \cite{wahba1990spline}. Its use in functional regression was formalized by Reiss and  Ogden \cite{reiss2009smoothing}. By contrast, Eilers and Marx  \cite{eilers1996flexible} introduced a difference penalty on spline coefficients, which is conceptually related but distinct. In this work, we adopt the former approach.
The considered penalty term in our work is equal to  \begin{eqnarray}\label{eq5}
 P(\lambda, \boldsymbol{b}) :=  \lambda  \int  \left(  \frac{\partial^2 \delta(t)}{\partial t^2} \right)^2 dt 
   = \lambda \boldsymbol{ b}^{\top} \boldsymbol{P_{1p}} \boldsymbol{ b},
\end{eqnarray}
where $\boldsymbol{P_{1p}}$ is the $L \times L$ positive definite matrix whose $(\ell, \ell')$ element is given by
$P_{1p}(\ell, \ell') = \int B_{\ell}^{\prime\prime}(t), B_{\ell'}^{\prime\prime}(t), dt$; these are fixed quantities, given that we use B-Spline basis. The parameter $\lambda$, known as the smoothing parameter, is a positive value that establishes a trade-off between the closeness of the fitted curve to the observed data and its smoothness.

\noindent For a given  $\rho$, the unknown coefficients $\boldsymbol{\theta}$ and $\lambda$ are estimated by fitting a linear mixed model, where the smoothing parameter is considered as a variance component \cite{ruppert2003semiparametric,reiss2009smoothing,wood2017generalized}, the penalized regression coefficients ($ \boldsymbol{b}$) are considered as random effects and all other unpenalized coefficients $(\zeta_0, \boldsymbol{\zeta, \alpha,\eta})$ are considered as fixed effects. The variance parameters are then estimated by ML/REML (maximum likelihood or restricted maximum likelihood).

\noindent In particular, we fit our proposed model by expressing it as the following linear mixed model
\begin{eqnarray}
    \boldsymbol{Y|b} \sim \boldsymbol{N}( \boldsymbol{X \beta }+  \boldsymbol{Z b} , \sigma^2 \boldsymbol{I}), \quad  \boldsymbol{ b} \sim \boldsymbol{N}\left(\boldsymbol{0}, \frac{\sigma^2}{\lambda}(\boldsymbol{P_{1p}})^{-1}\right),
\end{eqnarray}
with $\boldsymbol{X}=( \mathds{1}_N | \boldsymbol{W}|\boldsymbol{G}|\boldsymbol{K} )$ and $\boldsymbol{\beta}=(\zeta_0, \boldsymbol{\zeta, \alpha,\eta})^{\top}$.

\noindent The likelihood for $(\boldsymbol{b},\boldsymbol{\beta})$ is
$$ f(\boldsymbol{Y}|\boldsymbol{b},\boldsymbol{\beta}) = (2 \pi)^{- N/2} |\sigma^2 \boldsymbol{I}|^{-1/2} \exp \{ - ( \boldsymbol{Y} - \boldsymbol{X \beta } - \boldsymbol{Z b} )^{\top} (\sigma^2 \boldsymbol{I})^{-1} ( \boldsymbol{Y} - \boldsymbol{X \beta } - \boldsymbol{Z b} ) \} $$

$$ f(\boldsymbol{b}) = (2 \pi )^{-L/2} \vert (\sigma^2/ \lambda)(\boldsymbol{P_{1p}})^{-1} \vert^{-1/2} \exp \{ - \boldsymbol{b}^{\top} ((\sigma^2/ \lambda)(\boldsymbol{P_{1p}})^{-1}))^{-1}  \boldsymbol{b} ) \} $$

\noindent Then, we obtain 
\begin{align*}
    \log (f(\boldsymbol{Y,b}|\boldsymbol{\beta})) =
    &- \frac{N}{2} \log (2 \pi) - \frac{1}{2} \log (|\sigma^2 \boldsymbol{I}|) - \frac{L}{2} \log(2 \pi) - \frac{1}{2} \log (\vert (\sigma^2/ \lambda)(\boldsymbol{P_{1p}})^{-1} \vert)   \\
    &- \frac{1}{\sigma^2} (\boldsymbol{Y} - \boldsymbol{X} \boldsymbol{\beta} - \boldsymbol{Z} \boldsymbol{ b} )^{\top} (\boldsymbol{Y} - \boldsymbol{X} \boldsymbol{\beta} - \boldsymbol{Z} \boldsymbol{ b}  ) - \frac{\lambda}{\sigma^2}  \boldsymbol{ b}^{\top} \boldsymbol{P_{1p}} \boldsymbol{ b}.    
\end{align*}
Maximizing the likelihood  with respect to the unknown $(\boldsymbol{\beta}, \boldsymbol{b})$ leads to the minimization of 
\begin{eqnarray}
    \frac{1}{\sigma^2} (\boldsymbol{Y} - \boldsymbol{X} \boldsymbol{\beta} - \boldsymbol{Z} \boldsymbol{ b} )^{\top} (\boldsymbol{Y} - \boldsymbol{X} \boldsymbol{\beta} - \boldsymbol{Z} \boldsymbol{ b}  ) + \frac{\lambda}{\sigma^2}  \boldsymbol{ b}^{\top} \boldsymbol{P_{1p}} \boldsymbol{ b}.  
\end{eqnarray}
This shows that BLUP (Best Linear Unbiased Predictor) estimation of $(\boldsymbol{\beta}, \boldsymbol{b})$ implies generalized least squares with a penalty term (see Section 4.5.3 in Ruppert, Wand, and Carroll \cite{ruppert2003semiparametric}). \\
 The BLUP estimation of $\boldsymbol{\beta}$ and $\boldsymbol{b}$ is given by  

\begin{eqnarray}
   \boldsymbol{\hat{\theta}} = \begin{bmatrix}
    \boldsymbol{\hat{\beta}}  \\ \boldsymbol{\hat{b} } \end{bmatrix} = \left( \boldsymbol{ {A}^{\top} } (\sigma^2 \boldsymbol{I})^{-1} \boldsymbol{A} + \boldsymbol{S}_{\lambda,\sigma^2}\right)^{-1} \boldsymbol{ A^{\top}} (\sigma^2 \boldsymbol{I})^{-1} \boldsymbol{Y}  
\end{eqnarray}
with $ \boldsymbol{S}_{\lambda,\sigma^2} $ is a  block diagonal  matrix of dimension  $ (S + 2D + L +1) \times (S + 2D + L +1) $ defined as $ \boldsymbol{S}_{\lambda,\sigma^2} = \text{Diag} \{ 0, \boldsymbol{0_S}, \boldsymbol{0_{2D}},\frac{\lambda}{\sigma^2}\boldsymbol{P_{1p}}\} $. The variance of $\boldsymbol{Y}$ is given as
 $Var(\boldsymbol{Y})  = \sigma^2 \left(\boldsymbol{I} + \lambda^{-1}  \boldsymbol{Z} (\boldsymbol{P_{1p}}) \boldsymbol{Z}^{\top}   \right) = \sigma^2 \boldsymbol{V_{\lambda}}$, with $\boldsymbol{V_{\lambda}} = \left(\boldsymbol{I} +\lambda^{-1}  \boldsymbol{Z} (\boldsymbol{P_{1p}}) \boldsymbol{Z}^{\top}   \right ) $. The use of the mixed model allows us to estimate $\lambda$ and $\sigma^2$ as variance components by maximizing the following log-likelihood 

\begin{eqnarray}
    l(\boldsymbol{\theta},\lambda, \sigma | \boldsymbol{Y}) = - \frac{1}{2}  \{ \log|\sigma^2 \boldsymbol{V}_{\lambda}| + (\boldsymbol{Y} - \boldsymbol{A \theta} )^{\top} (\sigma^2 \boldsymbol{V_{\lambda}})^{-1} (\boldsymbol{Y} - \boldsymbol{A \theta} )  \}.
\end{eqnarray}
The restricted log-likelihood is 

\begin{eqnarray}
    l_R(\boldsymbol{\hat{\theta}},\lambda, \sigma | \boldsymbol{Y}) = - \frac{1}{2}  \{ \log|\sigma^2 \boldsymbol{V}_{\lambda}| + (\boldsymbol{Y} - \boldsymbol{A \hat{\theta}} )^{\top} (\sigma^2 \boldsymbol{V_{\lambda}})^{-1} (\boldsymbol{Y} - \boldsymbol{A \hat{\theta}} )  + \log | \sigma^{-2} \boldsymbol{A^{\top} }  \boldsymbol{V_{\lambda}^{-1}} \boldsymbol{A}  |   \}.
\end{eqnarray} 
Maximizing the restricted log-likelihood $ l_R(\boldsymbol{\hat{\theta}},\lambda, \sigma )$ allows us to obtain the REML estimators $\hat{\lambda}_{REML}$ and $\hat{\sigma}^2_{REML}$. We then substitute $\lambda$ and $\sigma^2$ with their  REML estimators to compute $\boldsymbol{\hat{\theta}}$  generated by the REML method.

  \noindent REML corrects for the degrees of freedom used in estimating fixed effects, which can reduce bias in the estimation of variance components, including the smoothing parameter $\lambda$ \cite{ruppert2003semiparametric}. Reiss and Ogden \cite{reiss2009smoothing} show that, at finite sample sizes, generalized cross validation (GCV) is more likely to develop multiple minima and under-smooth, relative to REML, while  REML tends to provide a more balanced bias-variance trade-off. GCV can sometimes lead to overly smooth or overly rough estimates, whereas REML is generally more consistent in finding an optimal $\lambda$ that balances bias and variance \cite{wood2017generalized}. Wood \cite{wood2017generalized} presents smoothing parameter estimation using a Bayesian framework where the smoothing parameters have priors. This Bayesian treatment can lead to estimates that are closely related to those obtained by REML. The author demonstrates that under certain conditions, the Bayesian estimates of the smoothing parameters (obtained by maximizing the posterior distribution) are equivalent to the REML estimates. The covariance matrix of $\hat{\boldsymbol{\theta}}$ is  $\boldsymbol{\hat{V}_{\theta}} = \left(  \boldsymbol{A^{\top}} (\hat{\sigma}^2 \boldsymbol{I}) \boldsymbol{A}  + \boldsymbol{S}_{\hat{\lambda},\hat{\sigma}^2}  \right)^{-1} \hat{\sigma}^2 $. In this work, we use REML to select the smoothing parameter $\lambda$.


\subsection{Hypothesis test of interaction}
\label{sec2.6}

We want to test the significance of the interaction in the model. Formally, this requires testing
$$ H_0 : \eta_d = 0 \quad \forall d \qquad   \text{vs}  \qquad  H_1 : \eta_d \neq 0  \quad \text{for a certain } d . $$
Since $ \mathbb{E}(\boldsymbol{\hat{\theta}}) = 0 $ if $ \boldsymbol{\theta} =0 $, we have that $ \mathbb{E}(\hat{\boldsymbol{\eta}}) \approx 0 $ if $ \boldsymbol{\eta} = 0 $.  When  we extract the estimated covariance matrix $\boldsymbol{\hat{\Sigma}}^{-1}_{\hat{\eta}}$ of $\boldsymbol{\hat{\eta}}$ from the covariance matrix of $\boldsymbol{\hat{\theta}}$, we have under the null hypothesis $H_0$ : $\boldsymbol{\eta} = 0 $,
\begin{eqnarray}
\boldsymbol{\hat{\eta}} \sim N(0, \boldsymbol{\Sigma_{\hat{\eta}}} ) .
\end{eqnarray}
Given that $\boldsymbol{\eta}$ is a subvector of $\boldsymbol{\theta}$ containing only unpenalized coefficients, and since the estimation of the scale parameter $\sigma^2$ is involved in the fitting process, we propose a Wald-type statistic following the approach of Wood \cite{wood2017generalized} 
\begin{eqnarray*}
T_D = (\boldsymbol{\hat{\eta}}^{\top} \boldsymbol{\hat{\Sigma}^{-1}_{\hat{\eta}}} \boldsymbol{\hat{\eta}})/D.
\end{eqnarray*}
Under the null hypothesis, this Wald statistic $T_D$ asymptotically follows a  Fisher distribution with $D$ and $N-S-L-2D-1$ degrees of freedom. Therefore, we rejet the null hypothesis at level $\alpha$ if $T_D > F_{D,N-S-L-2D-1,1-\alpha}$, where $F_{\nu_1,\nu_2,\gamma}$ is the quantile of the distribution $F_{\nu_1,\nu_2}$ defined by $\gamma=P(F_{\nu_1,\nu_2}<F_{\nu_1,\nu_2,\gamma})$.

\section{Simulation}\label{sec3}

We conducted simulations under the null and the alternative hypotheses. In all cases, we set $S=1$, $\zeta_0 = 0$ and $\zeta_1 = 0.3$ and we simulated one covariate, $W_1 \sim \mathit{N}(0, 0.1^2)$. We generated $D = 20$ SNPs, assuming minor allelic frequencies $f \sim \text{Uniform}[0.05, 0.2]$. For each SNP, we simulated $R_{1d} \sim \text{Bernoulli}(f)$ and $R_{2d} \sim \text{Bernoulli}(f)$ and set $G_d = R_{1d} + R_{2d} \in \{0, 1, 2\}$ for $d = 1, \ldots, 20$. We fixed the parameters $\alpha_d$ ($d = 1, \ldots, 20$) as follows: $
\alpha_1 = 1.20$, $\alpha_2 = 1.00$, $\alpha_3 = 0.80$, $\alpha_4 = 0.50$, $\alpha_5 = 0.30$, $\alpha_6 = 0.90$, $\alpha_7 = 1.60$, $\alpha_8 = 1.30$, $\alpha_9 = 0.67$, $\alpha_{10} = 0.89$, $\alpha_{11} = 1.45$, $\alpha_{12} = 1.40$, $\alpha_{13} = 0.45$, $\alpha_{14} = 0.70$, $\alpha_{15} = 1.35$, $\alpha_{16} = 0.95$, $\alpha_{17} = 0.55$, $\alpha_{18} = 0.88$, $\alpha_{19} = 1.30$, $\alpha_{20} = 0.50$.
The genomic positions of the SNPs were simulated from a uniform distribution over $[0, 1]$.

\noindent To generate DNA methylation data, we considered 8 samples from B cells MethylC-seq data taken near the BLK gene located on chromosome 8 \cite{lakhal2017smoothed}. These samples are derived from a cohort of healthy individuals in Sweden. Data were sequenced on the Illumina HiSeq2000 system. We focused on methylation levels estimated within the genomic region $[11190000pb, 11460000pb]$, where DMRs were detected \cite{lakhal2017smoothed}.  The numbers of CpG sites in each sample are provided in Table \ref{tab_31}.
\begin{table}[H]
\caption{Number of CpG sites for each of the samples from B cell}
    \label{tab_31}
 \centering
    \newcommand{\minitab}[2][l]{\begin{tabular}{#1}#2\end{tabular}}
    \setlength{\tabcolsep}{3.5pt}
\begin{tabular}{|c|c|c|c|c|c|c|c|c|} 
\hline  Id  &   1 &2 &3 &4 & 5 & 6 & 7 & 8 \\ \hline
Sites CpG   & $23710$& $24248$& $23572$& $20230$&  $24596$ & $21444$&  $23462$& $23569$ \\\hline
\end{tabular}
\end{table}

\noindent To increase the size of the database, we replicated each of the 8 samples in the original dataset $n$ times. The genomic positions of the reads in the replicated samples are the same as those in the original sample used to replicate. The replicated values of the methylation levels $p^{\prime}_{ij}$ were obtained from the original values $p_{ij}$ as follows : 
\begin{enumerate}
    \item[$\bullet$] Generate $\vartheta_i$  from $N(0, \sigma_t^2)$.
    \item[$\bullet$] Set $\Tilde{p}_{ij} = logit(p_{ij}) + \vartheta_i$, where $p_{ij}$ is the DNA methylation level at the genomic position $t_{ij}$ in the original sample.
    \item[$\bullet$] The replicated DNA methylation level is $p^{\prime}_{ij} = logit^{-1}(\Tilde{p}_{ij})$.
\end{enumerate}

\noindent In total, we generated $8n$ samples following the procedure described above and we combined them with the original samples, resulting in a final database containing  $N=8(n+1)$ samples. Applying the kernel smoothing method described in Equation (\ref{eq21.1}), we derived DNA methylation curves for each individual as the genomic variable.

\noindent We set $\delta(t) = \cos(3 \pi t)$ to generate data. Finally, $Y_i$ was generated using Equation (\ref{eq22.3}), rewritten with the specified parameters:

\begin{eqnarray}
Y_i = 0.3 W_{i1} + \sum_{d=1}^{20} \alpha_d G_{id} + \int_0^1 \cos(3\pi t ) \Pi_i(t) dt + \sum_{d=1}^{20} \eta_d G_{id} \int_0^1 e^{-\rho|t - u_d| } \Pi_i(t) dt + \varepsilon_i ,
\end{eqnarray}
where we numerically approximated the integrals $\int_0^1 \cos(3 \pi t) \Pi_i(t) dt $ and $\int_0^1 e^{-\rho |t - u_d| } \Pi(t) dt $ by Riemann sums.\
We simulated $ \varepsilon_i \sim \mathit{N}(0,\sigma^2) $ such that,
$$ \sigma^2 = \frac{1}{10}Var \left(  0.3 W_{i1} +   \sum_{d=1}^{20} \alpha_d G_{id} + \int_0^1 \cos(3\pi t ) \Pi_i(t) dt + \sum_{d=1}^{20} \eta_d G_{id} \int_0^1 e^{- \rho |t - u_d| } \Pi_i(t) dt \right).$$

\noindent The sample size was set to $N\in \{200,400\}$. We used a cubic B-spline basis of 10 functions to estimate $\delta(t)$ and used the REML method to estimate the penalty parameter $\lambda$.

\subsection{Simulations under $H_0$}
\label{3.1}

To generate data under $H_0$ (absence of interaction), we set 
$\eta_d	=0$ for $d=1,…,20$. Under this scenario, we didn't have to specify $\psi_{\rho}$ nor $\rho$ to generate the data. However, the three parametric specifications, $\psi_{\rho}$, and the values of $\rho$ (0.1,1 and 8) were considered to perform the tests. In Figures 1 to 3, we present the quantile-quantile plots for the obtained p-values bases on 1000 simulations under different scenarios. These graphics clearly show that the proposed approach controls very well the type I error rate, regardless of the choice of $\psi_{\rho}$ and the value of $\rho$ used to perform the test.

\begin{figure}[H]  
  \centering  
  \includegraphics[width=16cm,height=15cm]{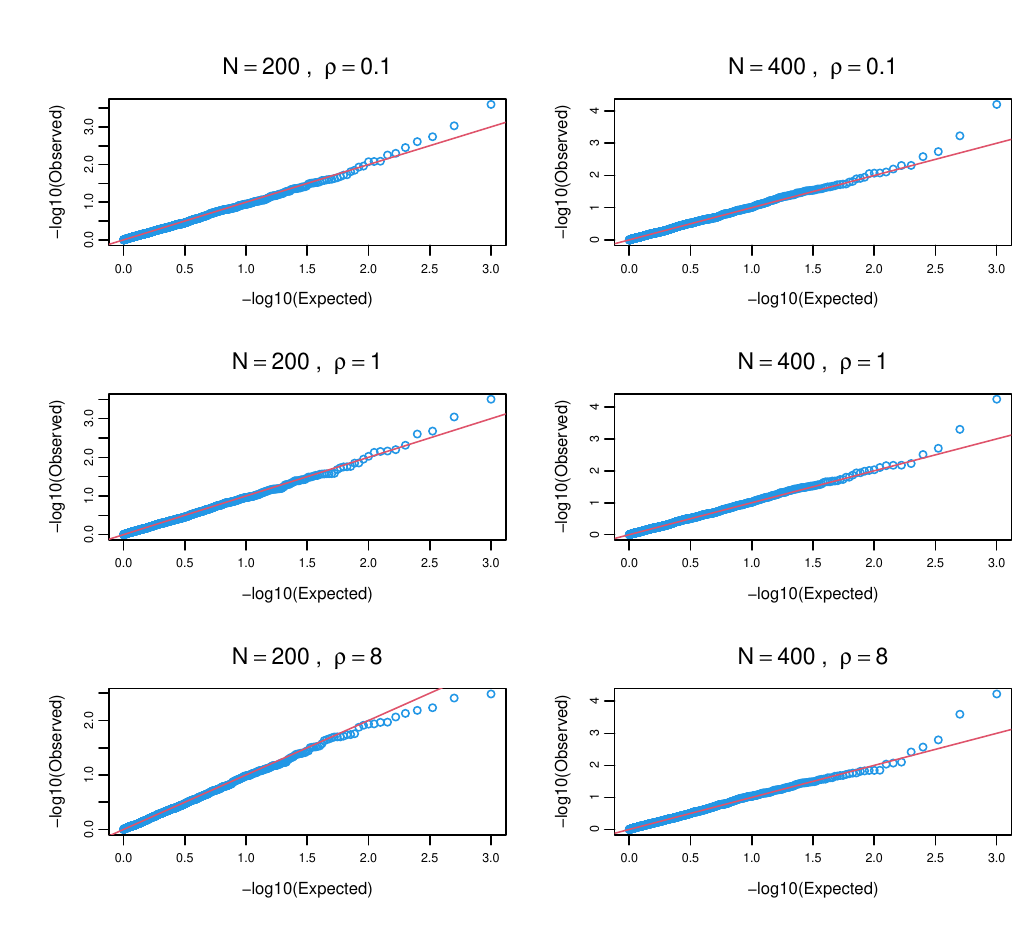}  
  \caption{Quantile-quantile plots of p-values with 1,000 simulations, data generated under $H_0$, model fitted with $\psi_{\rho}(u) = e^{- \rho u}$.} 
  \label{fig1}
\end{figure}

 \begin{figure}[H] 
  \centering  
  \includegraphics[width=16cm,height=17cm]{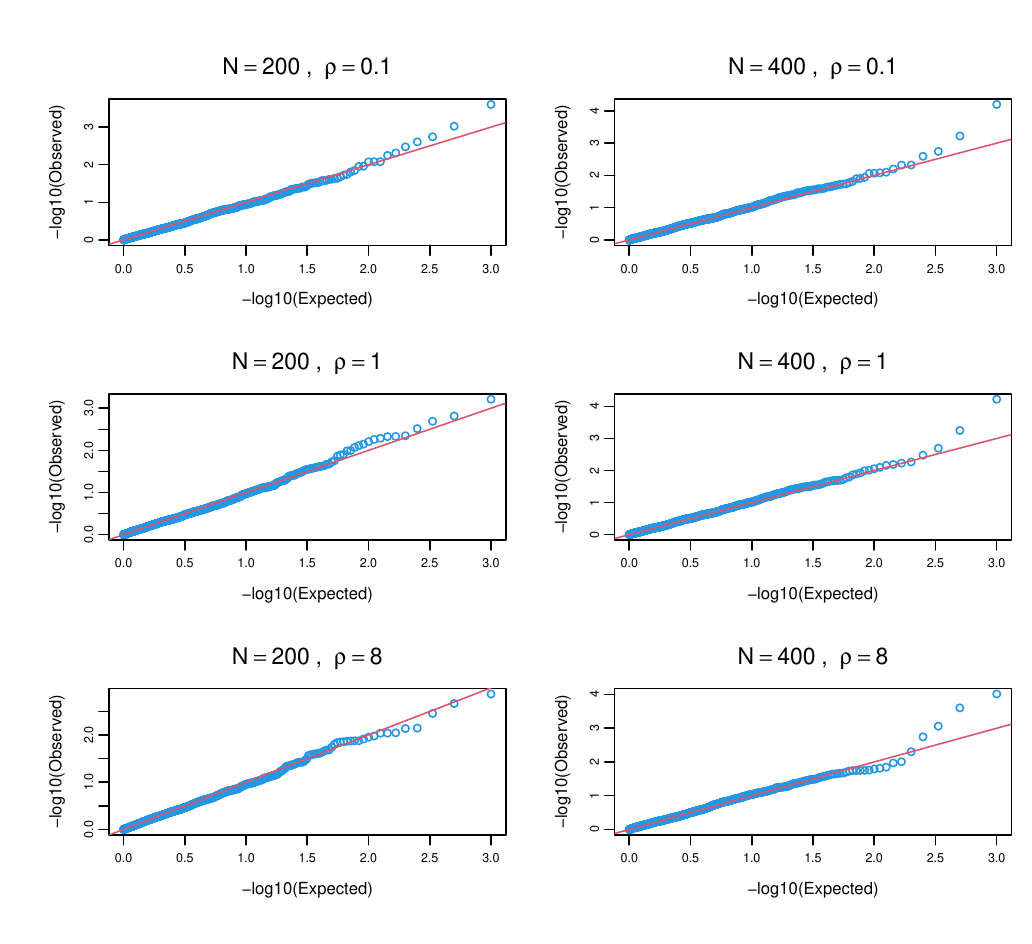}  
  \caption{Quantile-quantile plots of p-values with 1,000 simulations, data generated under $H_0$, model fitted with $\psi_{\rho}(u) = e^{- \rho^2 u^2}$.} 
   \label{fig2}
\end{figure}

\begin{figure}[H] 
  \centering  
  \includegraphics[width=16cm]{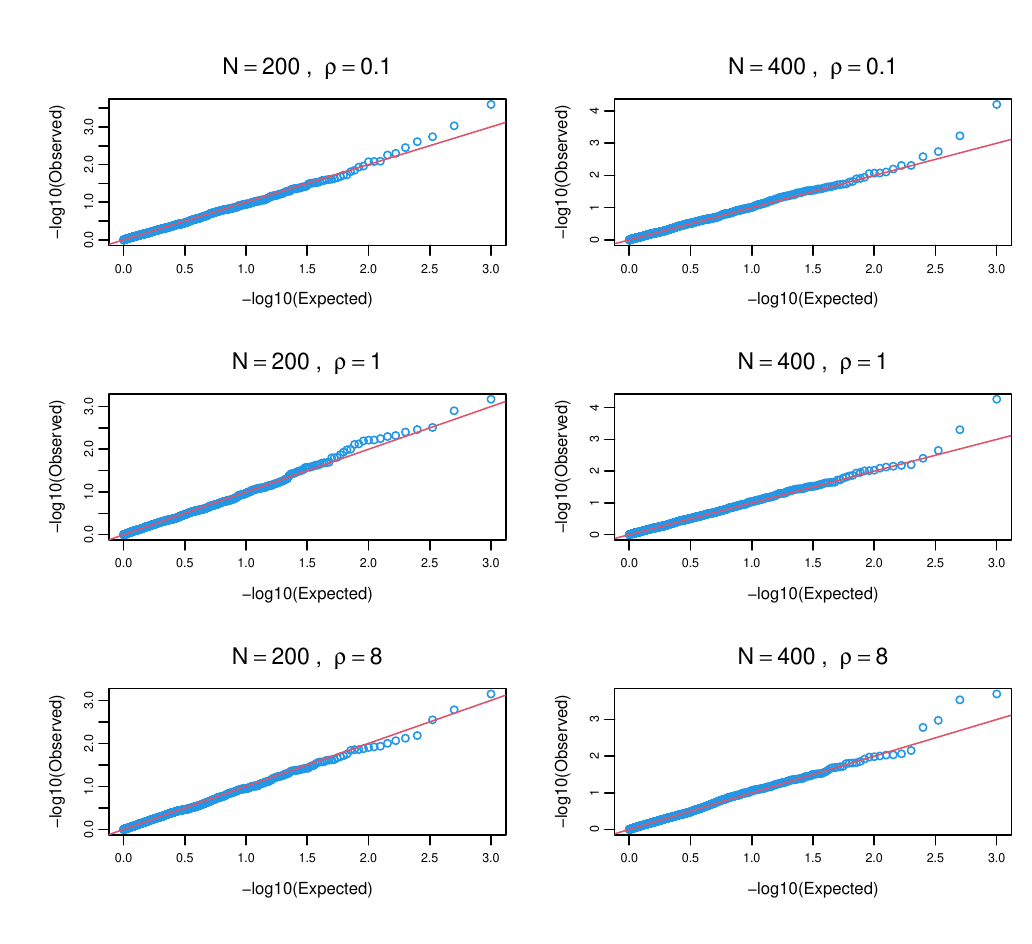}  
  \caption{Quantile-quantile plots of p-values with $1,000$ simulations, data generated under $H_0$, model fitted with $\psi_{\rho}(u) = \max(1- \rho u, 0) $.} 
   \label{fig3}
\end{figure}

\subsection{Simulations under $H_1$}
\label{3.2}

We conducted two simulation studies under the alternative hypothesis.

\begin{figure}[H]   
  \centering  
  \includegraphics[width=16cm]{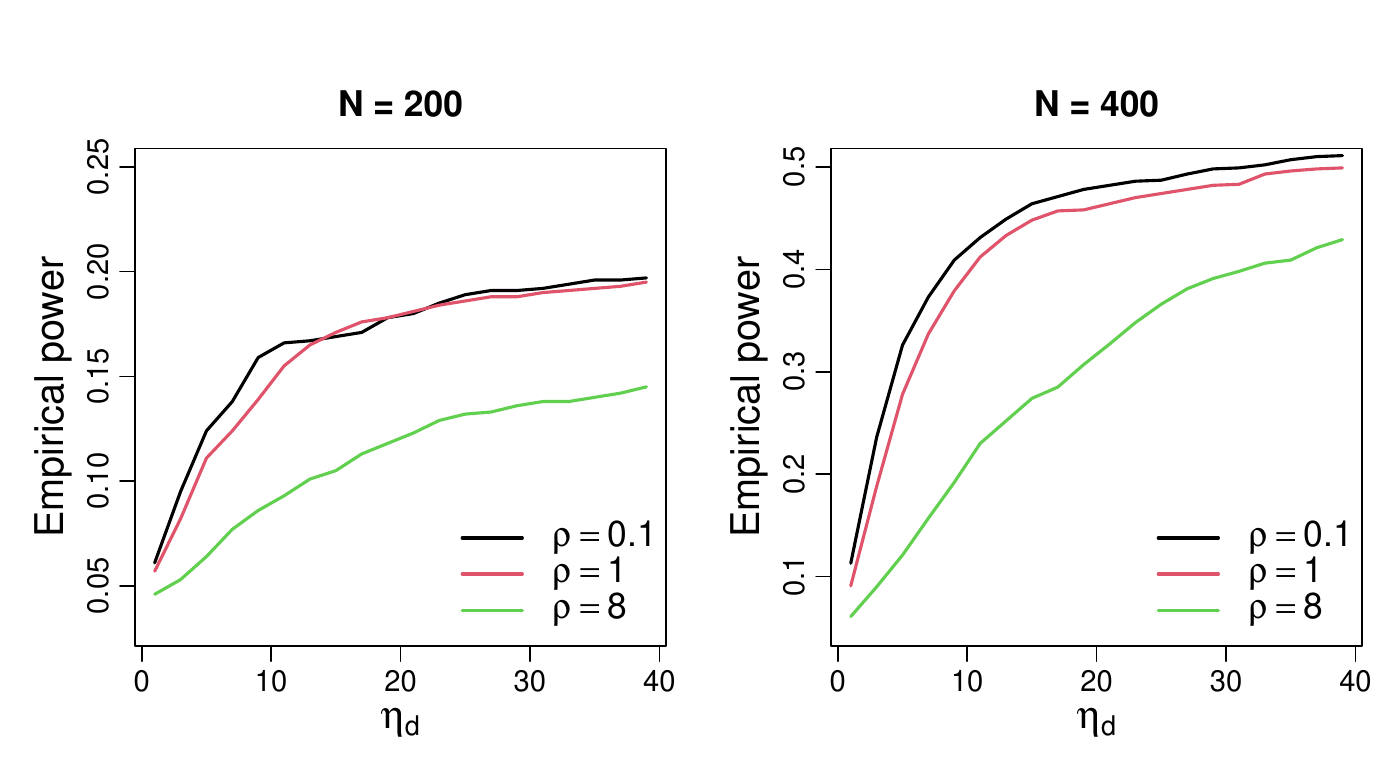}  
  \caption{Empircal power as a function of $\eta_d$ with correctly specified $\rho$ and $\psi_\rho$ from 1000 simulations with $N \in \{200,400\}$.} 
  \label{fig4}
\end{figure}

\noindent In the first study, we generated data under $H_1$ by specifying 3 values for $\rho$ (0.1,1 and 8) and by smoothly increasing the value of $\eta_d$ from 0 to 40. We performed the test by specifying the true function $\psi_\rho(u)=e^{-\rho u}$ with the true value of $\rho$. Figure \ref{fig4}, which show the empirical power, indicates that the power increases as the values of the interaction coefficients increase, regardless of the value of $N$. Additionally, the power increases when the value of $N$ increases and $\rho$ decreases, when the value of $\rho$ is correctly specified in the testing procedure.

\begin{figure}[H] 
  \centering  
  \includegraphics[width=16cm]{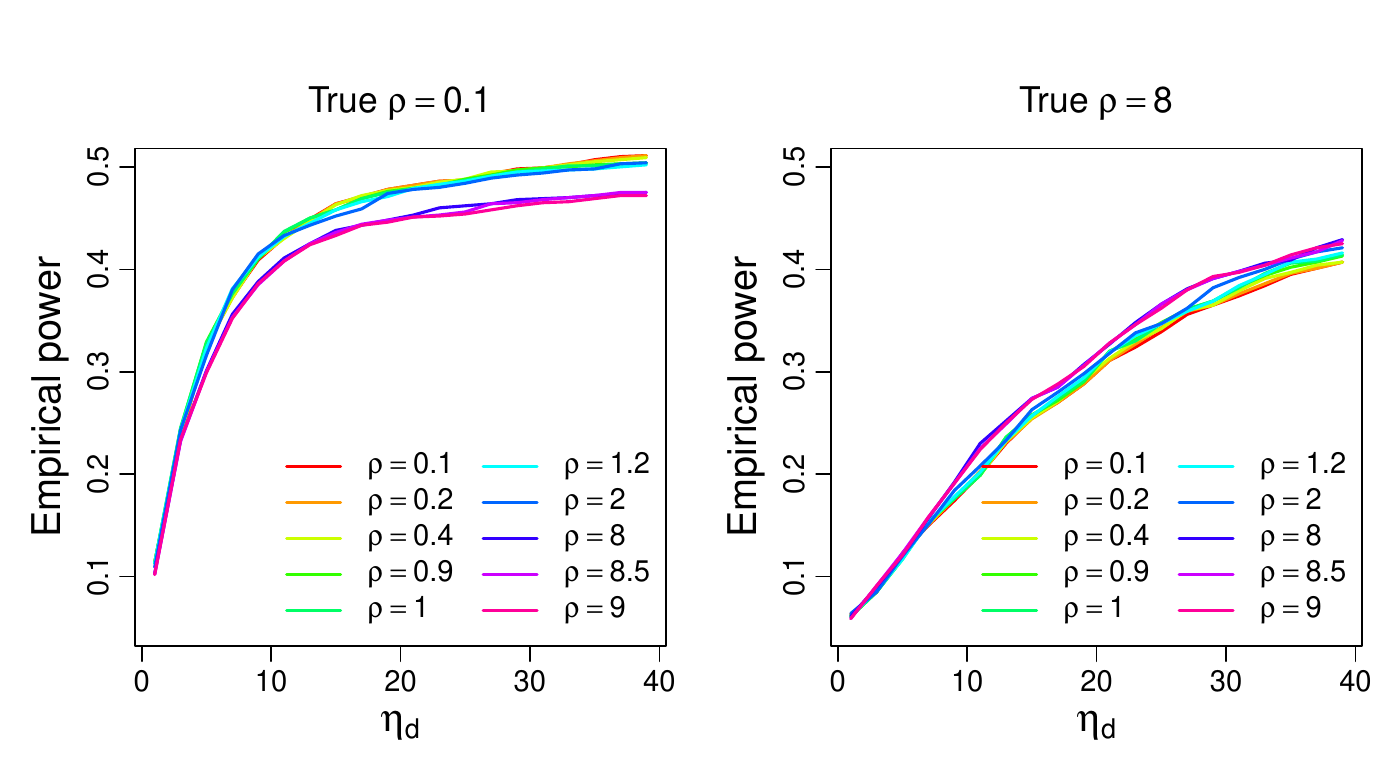 }  
  \caption{Empircal power as a function of $\eta_d$ with misspecified $\rho$ from 1000 simulations with $N= 400$. The true value of $\rho$ is 0.1 and 8 in the left handside and right handside graphics, respectively.} 
   \label{fig5} 
\end{figure}

\noindent The second study was aimed to investigate the impact of the misspecification of the value of $\rho$. We generated data using $\psi_\rho(u)=e^{-\rho u}$ with $\rho \in \{0.1,8\}$ and $N=400$ and performed the test using the same function $\psi_\rho(u)=e^{-\rho u}$ with $\rho \in \{0.1,0.2,0.4,0.9,1,1.2,2,8,8.5,9\}$. In Figure \ref{fig5}, we report the empirical power from 1000 simulations. From these graphics, we see the misspecification of $\rho$ has little impact on the empirical power. When the true $\rho$ is set to 0.1, we observed up to $5\%$ loss of power when we performed the test with $\rho \in \{8,8.5,9\}$. Similarly, we observed up to $3\%$ loss of power when the true value of $\rho$ is 9 and the test is performed with $\rho \in \{0.1,0.2,0.4,1,1.2,2\}$.

\noindent Finally, we observed a tiny impact on the empirical power when the parametric form of $\psi_\rho$ is misspecified.

\section{Analysis of real data}
\label{sec4}

We applied the proposed test to the obesity dataset GSE73103, described in Voisin et al. \cite{voisin2015many}. The original aim of their work  was to investigate how obesity-associated SNPs identified in GWAS might influence body mass through local DNA methylation changes, by testing SNP–CpG associations in blood samples from young individuals. The main goal of our study is to investigate the significance of the interactions between obesity-associated SNPs and DNA methylation. We also perform simulations based on this dataset to evaluate the performance of our method and compare it to the existing approaches \cite{veenstra2018epigenome,wang2022genetic}. 

\noindent This study involves data from $355$ healthy young individuals who were genotyped for $52$ known obesity-associated SNPs, previously identified in genome-wide association studies (GWAS) or meta-analyses as associated with obesity traits, and DNA methylation levels were obtained from their blood using the Illumina 450K BeadChip.  In the original study \cite{voisin2015many}, raw methylation data from the Illumina 450K array were preprocessed following standard procedures, including background correction (NOOB), probe-type bias adjustment (BMIQ), and batch-effect correction (ComBat). Probes on sex chromosomes, cross-reactive probes, and probes containing frequent SNPs were excluded. The resulting methylation values were expressed as $\beta$-values, representing the proportion of methylated alleles at each CpG site. In our analysis, we directly used these pre-processed $\beta$-values as the basis for constructing individual methylation curves. In their primary analysis, Voisin et al. \cite{voisin2015many} reported that 28 of the 52 obesity-associated SNPs were significantly associated with methylation levels at 107 nearby CpGs, many of which were located in promoters of genes previously implicated in obesity (e.g., BDNF, POMC, ADCY3), thereby highlighting the relevance of SNP–methylation interactions in obesity biology architecture. \\
The dataset comprises two sub-groups of healthy young Caucasians from two different age ranges: the first sub-group consists of $130$ individuals aged $14-16$ years, and the second sub-group consists of $225$ individuals aged $18-34$ years. We have $214$ males and $141$ females. Of these $355$ subjects, in the first sub-group, $101$ have normal weight, $6$ are obese, and $23$ are overweight, whereas in the second sub-group, $167$ have normal weight, $11$ are obese, and $47$ are overweight. We do not have acess to the continuous variable Body Mass Index (BMI) in the dataset, but rather the categorical variable Weight  with levels normal-weight, obese, and overweight.

\begin{figure}[t]  
  \centering  
  \includegraphics[width=16cm]{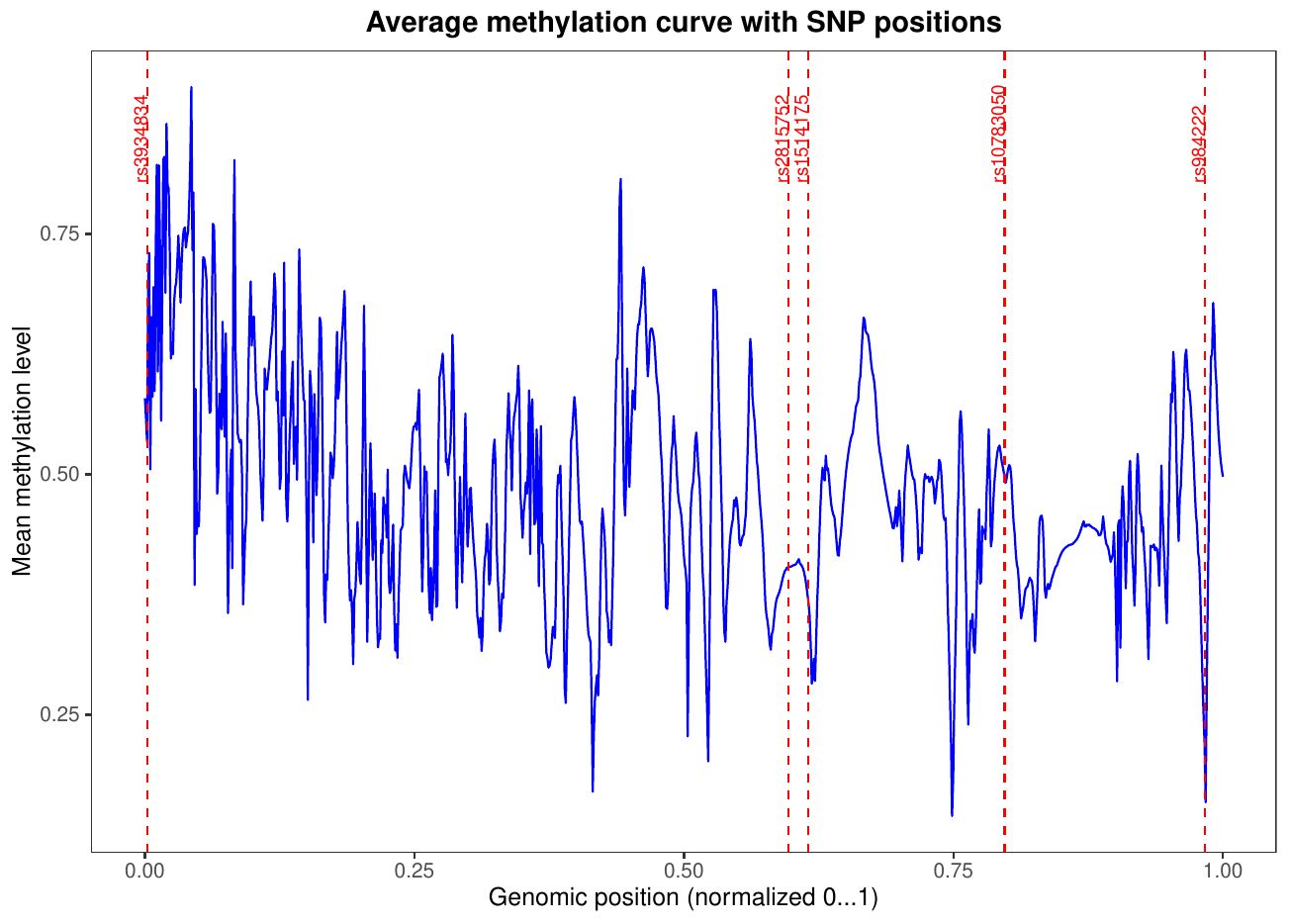}  
  \caption{Average methylation curve across 355 individuals in region 1, with vertical dashed lines marking the positions of the five obesity-associated SNPs (rs984222, rs2815752, rs3934834, rs1514175, rs10783050).}
  \label{fig:mean_methyl_curve}
\end{figure}

\noindent We created a binary variable coded as $0$ for individuals with normal weight and coded as $1$ for individuals who are obese or overweight. In our analysis, we focus on chromosome 1, where we have DNA methylation measurements for 39,869 CpG sites. We observed that there are no methylation measurements between genomic positions $121,485,060$bp and $142,618,825$bp. Therefore, we divided chromosome 1 into two regions: region1 contains genomic positions less than 1$21,485,060$bp, and region2 contains genomic positions greater than $142,618,825$bp. We have $5$  known obesity-associated SNPs in region1 (rs$984222$, rs$2815752$, rs$3934834$, rs$1514175$, rs$10783050$) and $2$  known obesity-associated SNPs in region2 (rs$516636$, rs$1011731$).  Subsequently, our analysis focuses on region 1, which includes DNA methylation measurements from $24,151$ CpG sites. We transform DNA methylation data into curves for each individual using Equation (\ref{eq21.1}). These curves serve as our genomic variables.\\
\noindent To provide a descriptive overview of the methylation landscape in region 1, Figure \ref{fig:mean_methyl_curve} shows the average methylation curve across the 355 individuals, together with the genomic positions of the five obesity-associated SNPs (rs984222, rs2815752, rs3934834, rs1514175, rs10783050). This visualization highlights the relative location of SNPs with respect to CpG sites and illustrates the variability of methylation levels along the genomic region.

\noindent Because differences in cell-type proportions between DNA samples can confound association results \cite{liu2013epigenome}, we followed the approach of Voisin et al.  \cite{voisin2015many}  by adjusting our analyses using a surrogate for cell-type proportions derived from 43 differentially methylated CpG sites present on the HumanMethylation450 array, which have the ability to discriminate between blood cell types \cite{koestler2012peripheral}. As a surrogate for cell-type proportions and to reduce the number of variables, we utilized the first two principal components (pca1, pca2) associated with these 43 sites, which together explain over $70\%$ of the total variance in methylation at these CpG sites, similar to Voisin et al. \cite{voisin2015many}. 

\noindent Firstly, we fitted a logistic regression model where the response variable is the binary variable created, and the explanatory variables are pca1 and pca2. Secondly, we fitted our proposed model  with the working residuals from the logistic model as continuous response variable. 

\begin{figure}[tb]  
  \centering  
  \includegraphics[width=13cm]{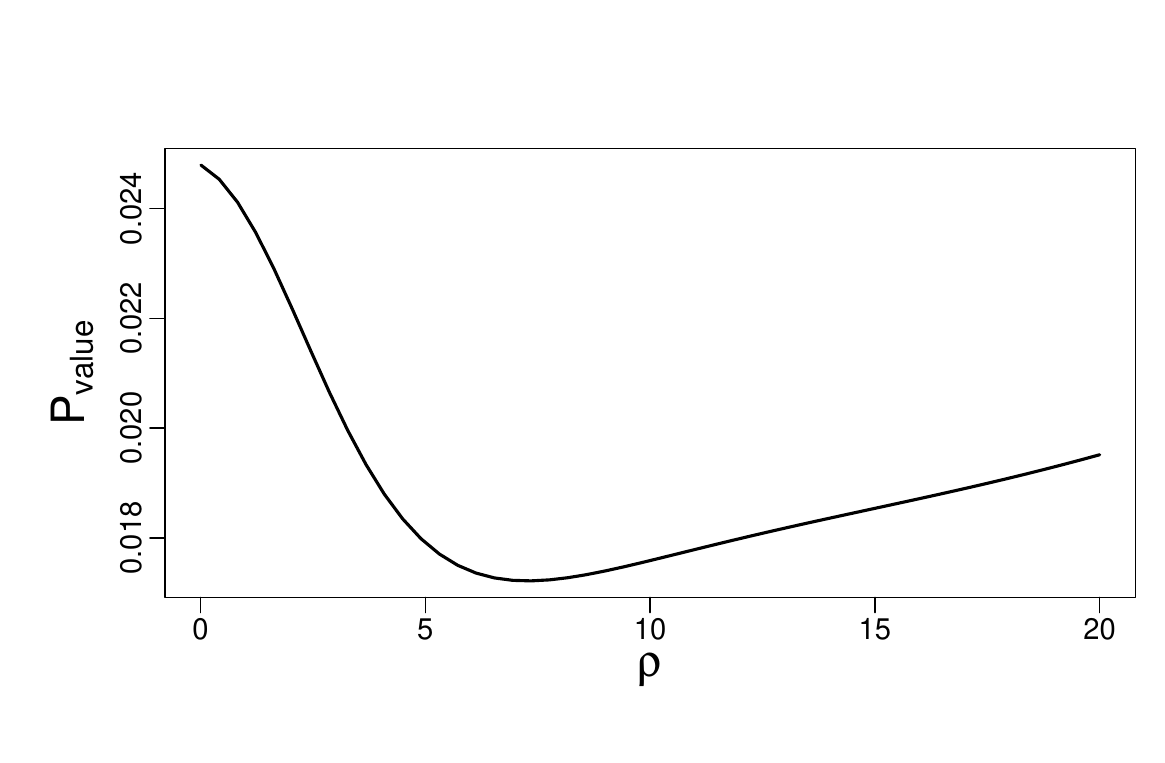}  
  \caption{Obtained p-values for the analysis of real data, as a function of the value of $\rho$ used to perform the test using  $\psi_\rho(u)=e^{- \rho u }$.} 
  \label{fig8}
\end{figure}

\begin{figure}[tb]  
  \centering  
  \includegraphics[width=13cm]{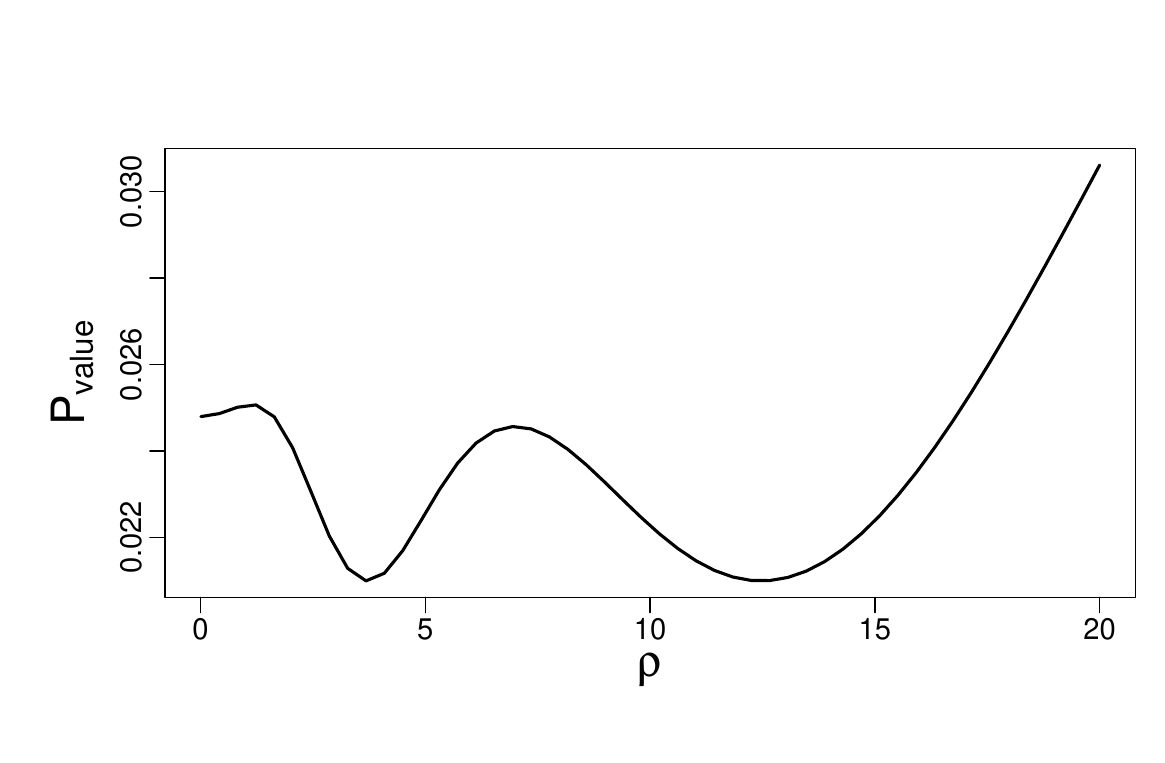}  
  \caption{Obtained p-values for the analysis of real data, as a function of the value of $\rho$ used to perform the test using  $\psi_\rho(u)=e^{- \rho^2 u^2 }$.} 
  \label{fig9}
\end{figure}

\begin{figure}[htbp!]  
  \centering  
  \includegraphics[width=13cm]{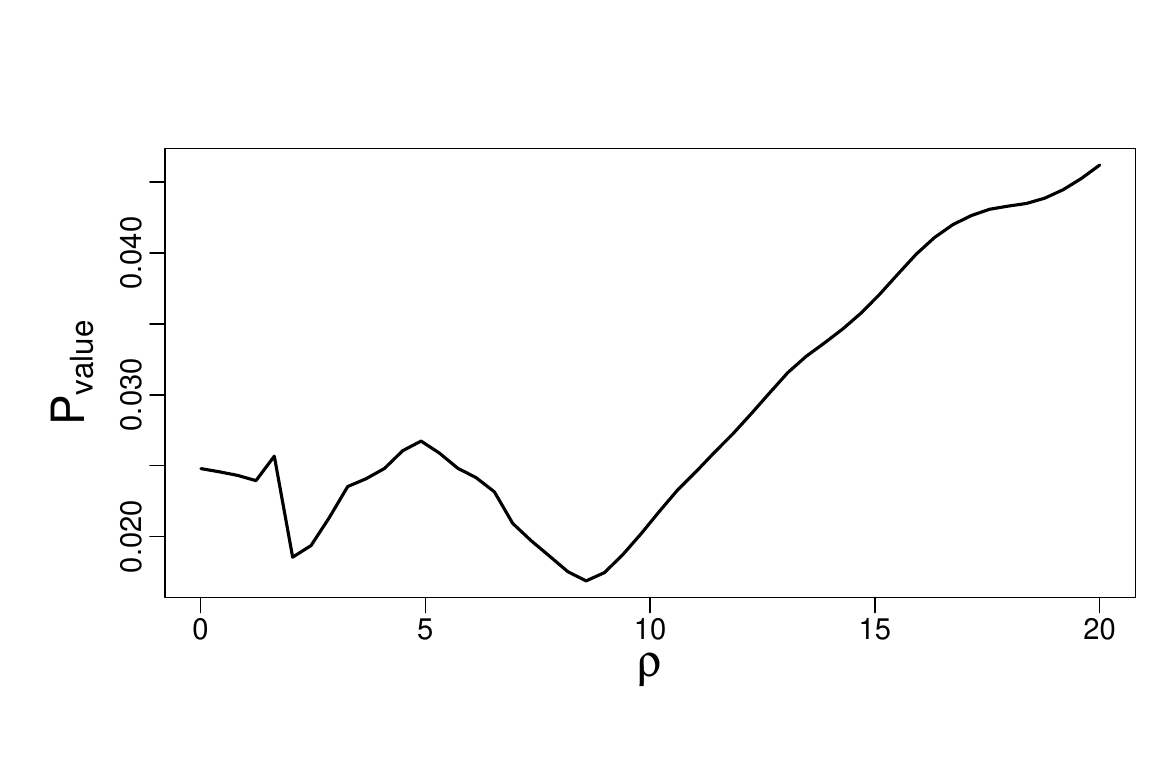}  
  \caption{Obtained p-values for the analysis of real data, as a function of the value of $\rho$ used to perform the test using  $\psi_\rho(u)=\max (1- \rho u , 0)$.} 
  \label{fig10}
\end{figure}

\noindent We used age and sex as covariates, methylation curves as the genomic variable, and SNPs as genetic variables. We performed the proposed test using for each of the parametric forms of $\psi_\rho$ with values of $\rho$ ranging from 0.1 to 20. In Figures \ref{fig8}  to \ref{fig10}, we reported the obtained p-values for each parametric form of $\psi_\rho$ as a function of $\rho$. These figures highlight a significant interaction regardless of the parametric form of $\psi_\rho$ and the value of $\rho$.

\begin{figure}[htbp!]  
  \centering  
  \includegraphics[width=10cm,height=8cm]{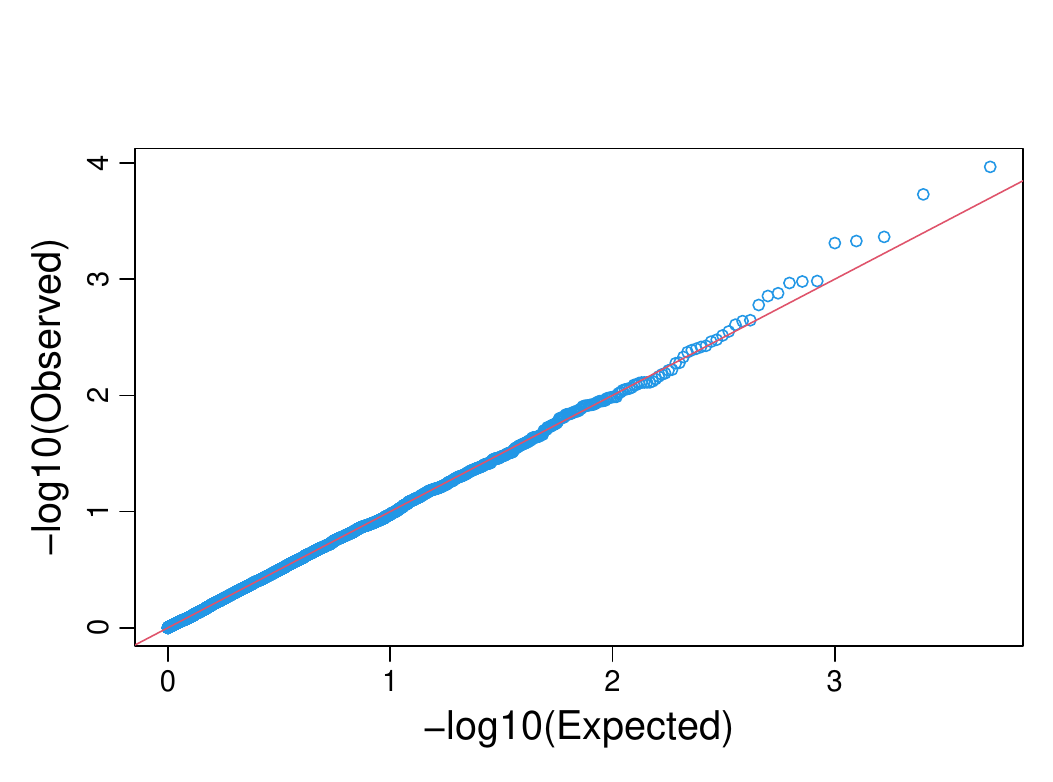}  
  \caption{Quantile-quantile plot of p-values from $5000$ simulations, data generated under $H_0$, model fitted with $\psi_{\rho}(u) = e^{- \rho u}$ and $\rho=10$, $N=355$.} 
  \label{fig6}
\end{figure}

\noindent We noticed, when fitting the model to the real data, that the residuals followed a mixture of two normal distributions: one with a mean of $-1.256$ and a variance of $0.2559$, with a proportion of $75\%$, and the other with a mean of $3.815$ and a variance of $0.4684$, with a proportion of $25\%$. We therefore conducted simulations in the absence of interaction (under H0) using real data to verify if the type I error is well controlled when the error distribution follows a mixture of normal distributions rather than a sigle normal distribution. To do this, as before, we first performed the test with $\psi_{\rho}(u) = e^{- \rho u}$ and $\rho=10$. Then, we used the estimated coefficients from the model as the true parameters to simulate the response variable $Y$ according to  Equation \ref{eq22.3}, while using an error term that is a mixture of normal distributions with the same specifications defined earlier. In Figure \ref{fig6}, we report the quantile-quantile plots for the obtained p-values from $5,000$ simulations. This Figure clearly shows the type I error rate is well controlled with the error terms are distributed following a mixte of normal distributions. 

\noindent Moreover, we conducted additional simulations using real data to compare the performance of our method to the existing method \cite{veenstra2018epigenome,wang2022genetic}, which we have named LintestSNPCpG. Five scenarios were considered. The SNP rs$3934834$ was used exclusively in the five scenarios. In the first scenario, we had interaction between one SNP and one CpG; in the second scenario, interactions between one SNP and $10$ CpGs; in the third scenario, interactions between one SNP and $20$ CpGs; in the fourth scenario, interactions between one SNP and $50$ CpGs; and in the fifth scenario, interactions between one SNP and $100$ CpGs. The SNP rs3934834 was used exclusively in the five scenarios. To generate data in the presence of interaction, we considered a simple linear model equation,  where $Y$ is the response variable and age, sex, the SNP, and the CpGs (with the number of CpGs varying from one scenario to another) are the explanatory variables, including interaction terms between the SNP and each of the CpGs depending on the scenario. We assumed that the error follows a mixture of two normal distributions as previously described.  To generated the response variable under scenario $1$, the true value of the intercept was set at $5$, the true value of the coefficient associated with the effect of age was set at $0.0798$, the coefficient associated with sex at $0.1521$, the coefficient associated with SNP at $-3$, and the coefficient associated with CpG at $-5.35744$. For scenario $2$, the true value of the intercept was set at $5.16$, $-0.5$ for the coefficient associated with SNP, $0.078$ for age, $0.225$ for sex, and $-0.2 $ for each of the CpGs. For the remaining three scenarios, we used the same values for the true coefficients: $0.2341$ for the intercept, $0.07203$ for age, $0.237$ for sex, $-0.5$ for each of the CpGs depending on the scenario, and $-0.2$ for the effect of SNP.

\noindent To adjust the existing model, we selected CpGs within a $500$ kb radius of the SNP, following the approach of Voisin et al. \cite{voisin2015many}. We detected $1080$ CpGs within this distance. We applied a Bonferroni correction, using a threshold of $0.05\%/1080$. For our proposed model, we used $10$ basis functions to capture the effect of methylation on the response variable and used the kernel $e^{-\rho u}$. We set $\rho=10$, a large value, to search for a local interaction around the SNP. We performed $1,000$ simulations in each scenario to compare our method to existing methods in terms of empirical power.

\begin{figure}[bp!]  
  \centering  
  \includegraphics[width=16cm]{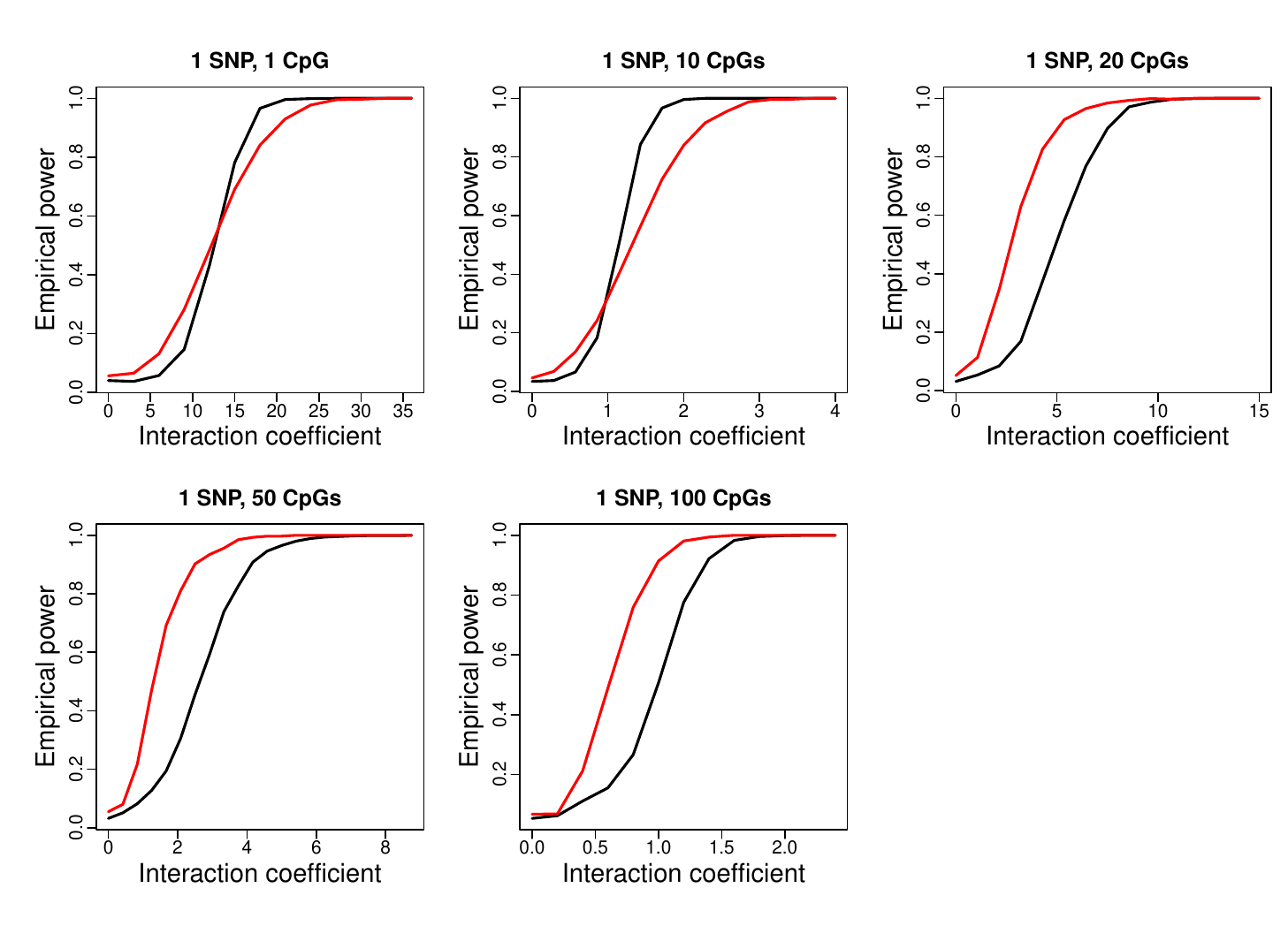}  
  \caption{Empirical power as a function of the value of the interaction coefficient, data generated in the presence of interaction, N=355. In red, we have the curve for proposed method, and in black, we have the curve for the existing method.} 
  \label{fig7}
\end{figure}

\noindent Figure \ref{fig7}  shows the power of the two methods for detecting interaction between the SNPs and CpGs under the five scenarios. It reveals that the proposed method outperformances the existing method in scenarios 3 to 5.


\section{Computational aspects and reproducibility}

All functions used to fit the proposed model to the real data are implemented in the R package \texttt{funInterMethSNP}, available on GitHub at 
\url{https://github.com/yvelingans/funInterMethSNP}.
The package provides vignettes that illustrate  the construction of methylation curves, the fitting of the model, and the interaction test used in this section.\\
\noindent The scripts used to generate all simulation scenarios (both under \(H_0\) and \(H_1\)), produce the QQ-plots and power curves are hosted in the repository \url{https://github.com/yvelingans/funInterMethSNP_SimuCont}.
The repository includes configuration files for each scenario and fixed random seeds to facilitate exact replication of the results reported in Section~\ref{sec3}.\\
 Transformation of methylation data into curves is a prerequisite for our method. For example, with the real dataset comprising approximately 24,151 CpG sites and 355 individuals, reconstructing methylation curves takes about 37.88 minutes on a MacBook Pro (Apple M2, 8 GB RAM). In contrast, the fitting of the functional models is performed very efficiently, although computation time increases progressively with the number of SNPs included. All model fitting was carried out in \texttt{R} using our functions built on the \texttt{gamm} routine from the \texttt{mgcv} package, which provides highly optimized algorithms for generalized additive models with penalized splines. To quantify the computational burden, we performed 10 replications of the main simulation scenarios and recorded the average running time; results are summarized in Table~\ref{tab:comp_time}. These results indicate that our method remains computationally feasible in candidate-gene or regional analyses.

 \begin{table}[ht]
\centering
\caption{Average computation time (in seconds) for fitting the proposed model, based on 10 replications per scenario. Computations were performed in R using \texttt{funInterMethSNP} on a standard laptop (MacBook Pro, Apple M2, 8~GB RAM).}
\label{tab:comp_time}
\begin{tabular}{ccc}
\hline
$N$ & Number of SNPs ($D$) & Average time (s)  \\
\hline
\multirow{4}{*}{200} 
 & 20  & 1.41  \\
 & 40  & 1.96  \\
 & 60 & 2.46  \\
 & 80 & 3.24  \\
\hline
\multirow{4}{*}{400} 
 & 20  & 2.74  \\
 & 40  & 3.71  \\
 & 60 & 4.78 \\
 & 80 & 5.91 \\
\hline
\end{tabular}
\end{table}

\section{Discussion}
\label{sec5}

The functional testing method introduced in this work offers a flexible and powerful approach to detect interactions between DNA methylation and SNPs across genomic regions. By integrating DNA methylation as functional data and allowing interaction's extent controler parameter $\rho$, the proposed model addresses the complexities of genetic and epigenetic interplay in influencing phenotypic traits.

\noindent Using extensive simulations, we demonstrated that the model effectively controls type I error rates and exhibits increased empirical power with higher interaction coefficients and appropriate settings of the interaction's extent controler parameter $\rho$. The application of the proposed method to real data from obesity patients and controls further underscores its utility, revealing superior performance over existing methods, especially when multiple interactions are present between SNPs and DNA methylation at CpG sites.

\noindent The application of the proposed method requires the specification of a parametric form for $\psi_\rho$ and a value for $\rho$. However, the simulations study and the real data analysis presented above show that this is not a major drawback that limits the usefullness of the proposed test. Based on our findings, we recommand using a small value of $\rho$ if interest lies in detecting interactions over a large genomic region. At the opposite, a large value of $\rho$ allows one to maximize power to detect a local interaction. By providing these guideline to selection $\rho$, we aim to facilitate the application of the proposed method in diverse research settings, ultimately contributing to a more nuanced understanding of the genetic and epigenetic factors driving complex traits and diseases.

\noindent  \textbf{Practical challenges in real data analysis.} 
The analysis of the obesity dataset involved several practical difficulties, including the use of a categorical weight variable instead of a continuous BMI measure, the need to adjust for cell-type heterogeneity, and the presence of a large genomic gap on chromosome~1 without methylation data.
In the original study by Voisin et al. \cite{voisin2015many}, the authors explicitly chose to use weight categories rather than BMI because the dataset included participants both under and over 18 years of age, and BMI scales differ between adolescents and adults. As our model is designed for continuous outcomes, this limitation prevented a direct application using an original continuous response and motivated our two-step approach, where residuals from a logistic regression were used as continuous inputs in the functional model.
In addition, the residual distribution deviated from normality, which motivated additional simulations to verify type~I error control under mixture-normal errors.

\noindent \textbf{Number of parameters exceed sample size : } 
 Our method is feasible in typical candidate-gene or regional analyses. However, extending the approach to genome-wide applications would require pre-selection of genomic regions, since the number of parameters to be estimated should not exceed the sample size. This limitation is not prohibitive when analyzing targeted regions, but it calls for the development of strategies capable of incorporating a larger number of SNPs while avoiding excessive multiple testing. Filtering approaches based on prior biological knowledge, the selection of SNPs already identified as associated with the phenotypic trait, or new forms of regularization on the number of SNPs in addition to functional regularization could provide promising solutions.

\noindent  \textbf{Additional simulations and future directions :}
To further evaluate the robustness of the proposed method with increasing numbers of SNPs, we conducted additional simulations where the number of SNPs was progressively increased up to 100, with $N=400$ individuals. The results showed that the type I error remained well controlled, but the empirical power decreased compared to the case of 20 SNPs, reflecting the larger number of parameters relative to the sample size. Increasing $N$ increased power. As with other functional regression frameworks, however, the number of parameters cannot exceed the sample size $N$, which limits the direct applicability when the number of SNPs reaches thousands. In such cases, a practical solution is to partition the genome into regions and perform region-based analyses, akin to GWAS strategies. Future work will focus on incorporating penalization or dimension-reduction techniques to extend the method to high-dimensional contexts where the number of SNPs far exceeds the sample size, while maintaining interpretability and computational feasibility.

\noindent \textbf{Context within functional approaches in genomics.}
The proposed framework contributes to a growing body of literature that integrates functional data analysis (FDA) methods into genomics and epigenomics. Recent advances in scalar-on-function and function-on-scalar regression models \cite{goldsmith2011penalized, ivanescu2015penalized, crainiceanu2024functional} have shown the potential of representing omics signals as smooth functional predictors, enabling more interpretable inference and reduced dimensionality compared to site-by-site analyses. Our work extends these ideas by explicitly modeling the interaction between functional methylation profiles and discrete genetic variants. This offers an interpretable and computationally efficient alternative to standard pairwise testing procedures typically used in epigenome-wide studies.

 \addcontentsline{toc}{section}{Bibliographie} 


\bibliographystyle{ama}
\bibliography{sample}

\newpage
\section*{Appendix}

\addcontentsline{toc}{section}{Appendix}

\renewcommand{\thefigure}{A\arabic{figure}}
\setcounter{figure}{0}

\subsection*{Additional simulations with 100 SNPs}

\begin{figure}[H]
  \centering
  \includegraphics[width=16cm]{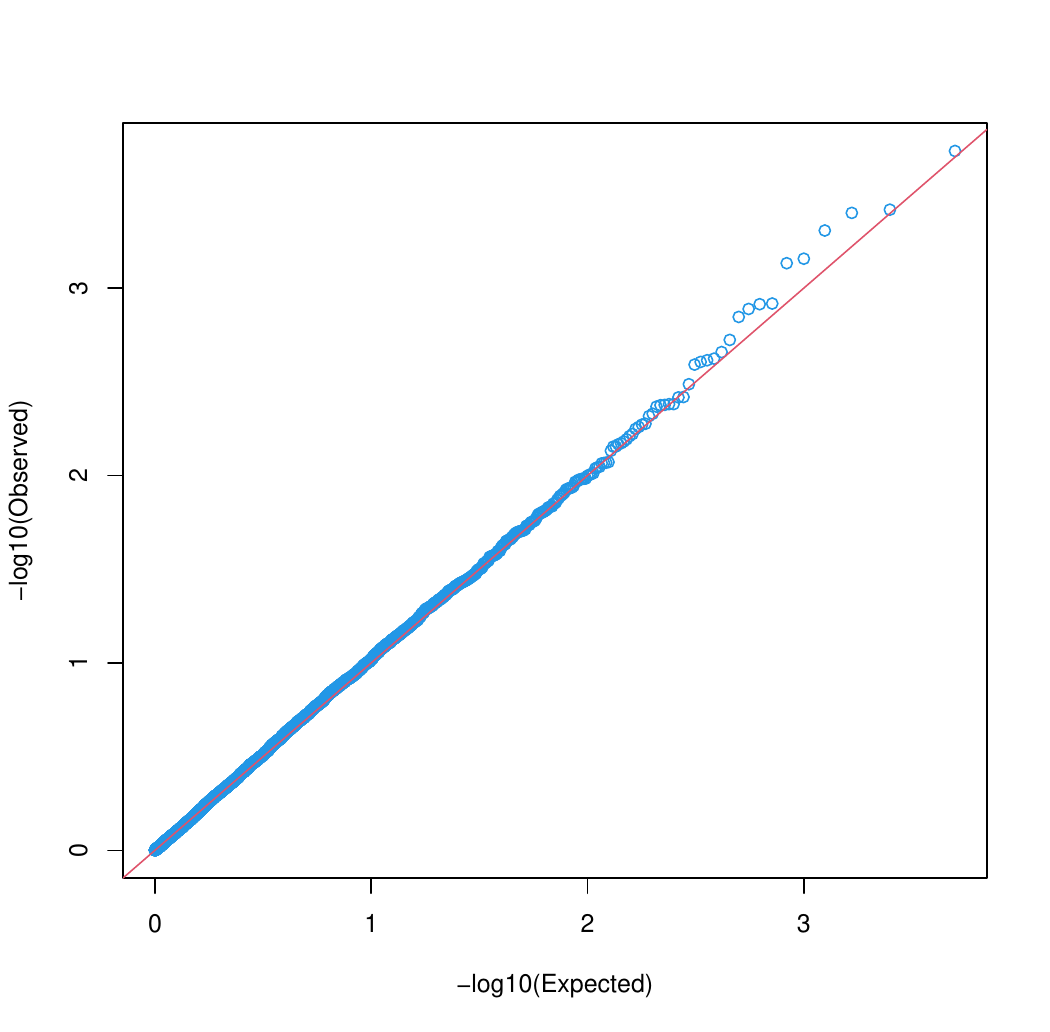}
  \caption{Quantile–quantile plot of p-values under $H_0$ with $D=100$ SNPs and $N=400$ individuals, 
  based on 5,000 simulations using $\psi_\rho(u)=e^{-\rho u}$ with $\rho=0.1$. 
  The alignment with the diagonal confirms that the type~I error rate remains well controlled even when the number of SNPs increases.}
  \label{fig:A1}
\end{figure}

\begin{figure}[H]
  \centering
  \includegraphics[width=16cm]{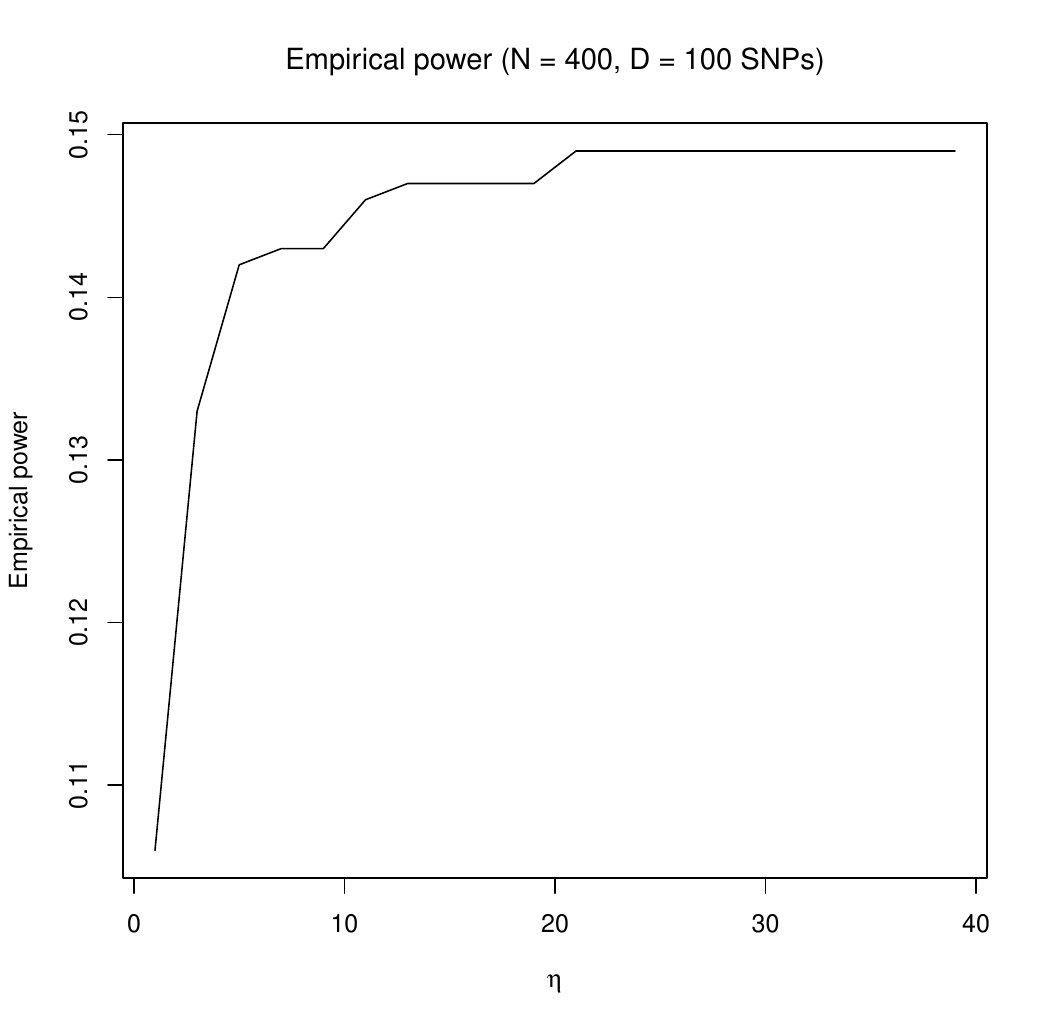}
  \caption{Empirical power of the proposed test with $D=100$ SNPs and $N=400$ individuals, averaged over 1,000 simulations. 
  The power increases slightly from 0.10 to 0.15 with growing interaction strength, then plateaus, reflecting the limited effective sample size relative to the number of SNPs.}
  \label{fig:A2}
\end{figure}

\begin{figure}[H]
  \centering
  \includegraphics[width=16cm]{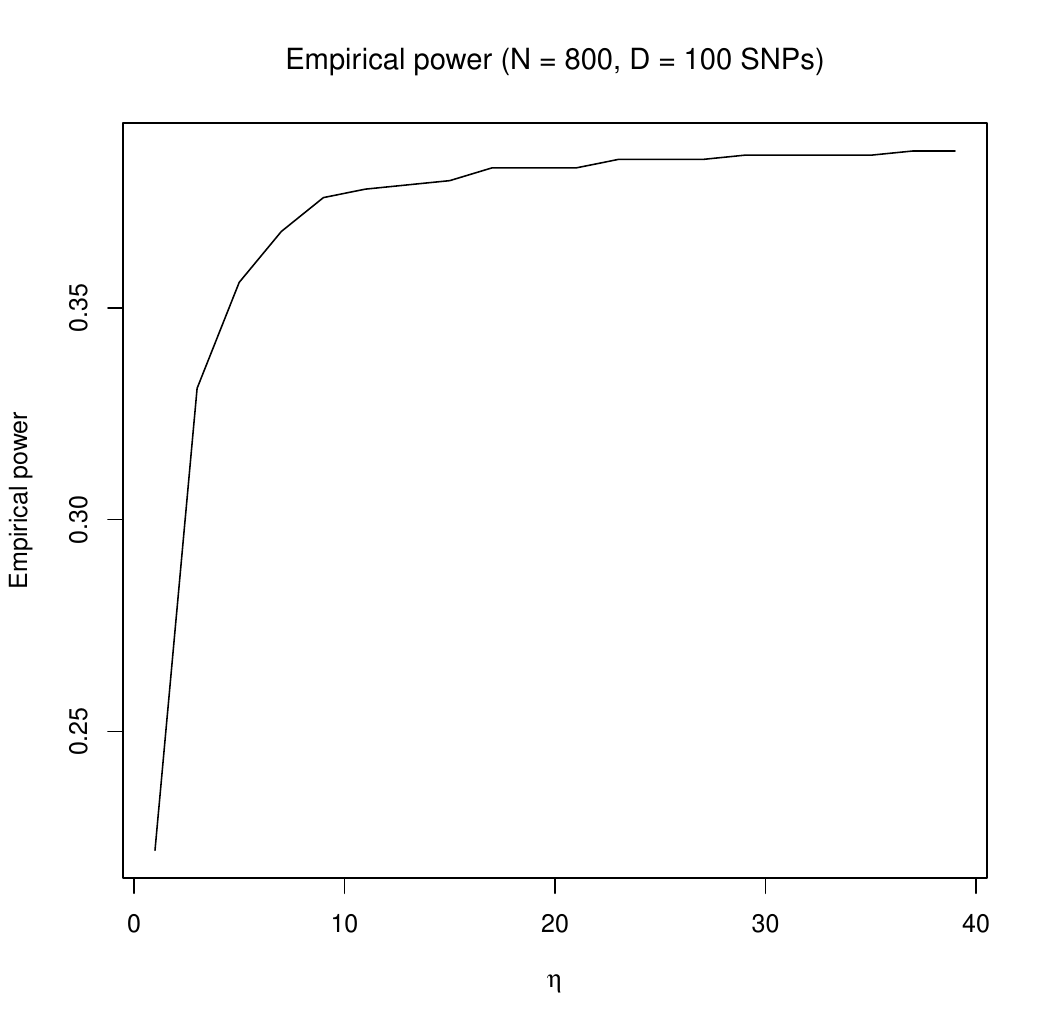}
  \caption{Empirical power of the proposed test with $D=100$ SNPs and $N=800$ individuals, averaged over 1,000 simulations. 
  The power rises from approximately 0.22 to 0.39 as the interaction strength increases, 
  demonstrating the expected improvement with larger sample sizes and confirming the scalability of the proposed method.}
  \label{fig:A3}
\end{figure}

\end{document}